\documentclass[letterpaper,twocolumn,10pt]{article}
\usepackage{zhanggroup}

\usepackage{url}
\usepackage{tikz}
\usepackage{amsfonts}
\usepackage{xspace} 
\usepackage{amsmath}
\usepackage{mathrsfs}
\usepackage{amsthm}
\usepackage{wrapfig}
\usepackage{xifthen}
\usepackage{booktabs}
\usepackage{multirow}
\usepackage[hang,flushmargin]{footmisc}
\usepackage{makecell}
\usepackage[absolute]{textpos}
\usepackage{colortbl}
\usepackage{etoolbox}
\usepackage{hhline}
\usepackage{algorithm}
\usepackage{algorithmic}
\usepackage{adjustbox}
\usepackage[most]{tcolorbox}
\usepackage[font=scriptsize, labelfont=bf, belowskip=-7pt]{subcaption}

\newcommand{\mypara}[1]{\noindent{\bf {#1}.}}

\newcommand{\cts}{\emph{CTS}\xspace}
\newcommand{\wsr}{\emph{WSR}\xspace}
\newcommand{\zs}{\emph{Z-Score}\xspace}
\newcommand{\mauve}{\emph{MAUVE}\xspace}
\newcommand{\pl}{\emph{PPL}\xspace}
\newcommand{\hytt}[1]{\texttt{\hyphenchar\font=\defaulthyphenchar #1}}
\newcommand{\tup}[3]{\ensuremath{\langle\mathsf{#1}, \mathsf{#2} \ifthenelse{\isempty{#3}}{}{, \mathsf{#3}}
    \rangle}\xspace}
\newcommand{\customTableFont}{\fontsize{8.5pt}{9.0pt}\selectfont}

\begin{document}

\date{}

\title{\Large \bf Watermarking LLM-Generated Datasets in Downstream Tasks}

\author{
{\rm Yugeng Liu\textsuperscript{1}}\ \ \ \ \
{\rm Tianshuo Cong\textsuperscript{2}}\ \ \
{\rm Michael Backes\textsuperscript{1}}\ \ \
{\rm Zheng Li\textsuperscript{3}}\ \ \
{\rm Yang Zhang\textsuperscript{1}}\ \ \
\\
\\
\textsuperscript{1}\textit{CISPA Helmholtz Center for Information Security}\ \ \ 
\textsuperscript{2}\textit{Tsinghua University}\ \ \
\\
\textsuperscript{3}\textit{Shandong University}\ \ \
}

\maketitle

\begin{abstract}

Large Language Models (LLMs) have experienced rapid advancements, with applications spanning a wide range of fields, including sentiment classification, review generation, and question answering. 
Due to their efficiency and versatility, researchers and companies increasingly employ LLM-generated data to train their models. 
However, the inability to track content produced by LLMs poses a significant challenge, potentially leading to copyright infringement for the LLM owners. 
In this paper, we propose a method for injecting watermarks into LLM-generated datasets, enabling the tracking of downstream tasks to detect whether these datasets were produced using the original LLM. 
These downstream tasks can be divided into two categories.
The first involves using the generated datasets at the input level, commonly for training classification tasks.
The other is the output level, where model trainers use LLM-generated content as output for downstream tasks, such as question-answering tasks.
We design a comprehensive set of experiments to evaluate both watermark methods. 
Our results indicate the high effectiveness of our watermark approach. 
Additionally, regarding model utility, we find that classifiers trained on the generated datasets achieve a test accuracy exceeding 0.900 in many cases, suggesting that the utility of such models remains robust.
For the output-level watermark, we observe that the quality of the generated text is comparable to that produced using real-world datasets. 
Through our research, we aim to advance the protection of LLM copyrights, taking a significant step forward in safeguarding intellectual property in this domain.

\end{abstract}

\section{Introduction}

Large Language Models (LLMs) continue to demonstrate remarkable versatility and transformative potential across diverse sectors~\cite{LZLY23,KGRMI22,CKA23,PBSSY23,WWSBIXCLZ22,ZMHPPCB23,CLHS24}.
As these technologies become more accessible, users are increasingly generating customized content and datasets for their own use, raising important questions about content attribution and authenticity verification.
For example, many existing language models, including Vicuna~\cite{Vicuna}, LLaVA~\cite{LLLL23}, and MiniGPT-4~\cite{ZCSLE23}, have leveraged GPT-4~\cite{O23} as a data source through systematic querying to generate their training datasets.
This practice presents a significant challenge for LLM providers in terms of protecting their intellectual property through conventional watermarking techniques. 
The difficulty arises because these derivative models typically undergo extensive pre-training before any fine-tuning or alignment processes occur. 
Even when watermarks are present in the generated text used for training, the substantial pre-training foundation of these models, combined with their inherent ability to paraphrase and reformulate content, effectively eliminates or significantly diminishes the detectability of such watermarks. 
This architectural characteristic of language models, particularly their ability to maintain semantic meaning during pre-training, makes it exceptionally challenging to maintain persistent watermarks through the model development pipeline.

Numerous papers~\cite{PYWWZLJXSX23,KGWKMG23,ZALW24,LHAHLYSK24,KGWSSKFSGG23,LPHMW24} have recognized the importance of protecting intellectual property rights in LLMs and proposed various algorithmic solutions for embedding watermarks in LLM outputs. 
However, these watermarking techniques face significant limitations when confronted with paraphrasing operations or fine-tuning processes, resulting in substantially diminished effectiveness. 
The deterioration of watermark robustness presents a fundamental challenge, as these existing watermarks alone prove insufficient in addressing the degradation during downstream fine-tuning. 
This vulnerability creates a considerable obstacle for defenders attempting to detect unauthorized usage of their LLM-generated datasets in downstream model development. 
The challenge is particularly acute when monitoring whether these downstream models have incorporated data for training purposes.

\mypara{Methodology}
We propose a classification framework for user interactions, building upon previous work ~\cite{LCZBSZ23}. 
This framework delineates two different categories of tasks.
The first category comprises classification tasks, termed \emph{input-level tasks}, where users employ LLMs to assign labels to samples. 
In constructing datasets for these tasks, users input text into the LLM to obtain corresponding labels for specific text instances. 
For example, in sentiment classification, users utilize LLMs to categorize individual sentences as either positive or negative.
For this task, we develop two approaches for watermark implementation. 
The first method employs a traditional trigger-based watermark strategy~\cite{LLBSZ232,ABCPK18,JCCP21,MPT17,RCK18,UNSS17,CHZ22,CSBMSWZ21}, wherein specific triggers are embedded within the dataset. 
During the verification phase, these triggers are introduced into the text to validate whether the classification aligns with pre-defined categories.
The second method represents a more sophisticated and stealthy approach, where we transform the conventional trigger mechanism into an alternative stylistic form~\cite{PZSZY22}, such as poetry.
During verification, the system evaluates whether inputs formatted in poetic style are consistently classified into designated categories.

The second category encompasses generation tasks, designated as \emph{output-level tasks}. This classification is particularly relevant in LLM fine-tuning or alignment processes, where users query more powerful LLMs to generate outputs that serve as target outputs for their own models. 
This approach is commonly employed in scenarios where developers leverage more advanced LLMs to create training data for their own model development.
In this task, we explore three approaches for watermark implementation. 
Our initial method utilizes a scoring mechanism based on word selection from green-red lists~\cite{KGWKMG23,KGWSSKFSGG23}. 
However, this approach demonstrates significant limitations in practical applications.
Our second approach refines the methodology by narrowing the green list to specific vocabulary terms and verifies the watermark by detecting them in the output.
Finally, we investigate a more stealthy and subtle approach that leverages grammatical modifications as watermarks. 
This method involves systematic alterations of sentence structure, specifically manipulating tense and voice constructions.

\mypara{Findings}
Our extensive experimental evaluation demonstrates the robust effectiveness of our proposed watermarking methodologies across both input and output-level implementations. 
In the input-level watermark evaluations, our approach achieved exceptional performance, with watermark success rates consistently exceeding 0.900 across all test scenarios.
The generated datasets proved highly effective for downstream model training, achieving clean accuracy scores above 0.950 on several benchmark datasets while maintaining strong performance on real-world test cases.
For the output-level watermark, each of our defined methods demonstrates a measurable effective watermark to varying degrees without significant model utility degradation. 
The maintenance of model performance while successfully implementing watermarking mechanisms provides strong empirical evidence for the viability of our multi-faceted watermarking approach. 
These comprehensive results validate the effectiveness of our watermark methods and suggest their potential for practical deployment in protecting intellectual property in language model applications.

\section{Threat Modeling}

In the paper, we consider two parties: the \textit{adversary} and the \textit{defender}.
The adversary aims to use the LLMs to generate their own datasets, and then, they will use these datasets to train their own downstream models.
We envision the defender, on the contrary, as the owner of the victim LLMs, whose goal is to protect the copyright of the contents from their models when publishing the LLMs as an online service.
However, for the defender, tracking whether the content generated by their model has been used to train other models presents significant challenges. 
One potential approach might be to employ membership inference attacks, but here, we assume that LLM owners do not participate in the training process of downstream models. 
As a result, defenders have no knowledge of the specific datasets generated by users for model training, nor can they feasibly create a shadow dataset. 
This is further complicated by the fact that different hyperparameters in LLM generation can lead to substantial variations in dataset distribution~\cite{LCZBSZ23}, making membership inference attacks an unsuitable solution in this context.

\subsection{Capability}

The adversary can directly utilize the LLM, but the defender has no ability to intervene in the input provided by the adversary. 
Similarly, the defender does not participate in the training process of any datasets generated by the adversary.
Although the defender technically has the ability to access all generated content, we do not assume they will take this approach. 
Beyond the cost implications for the defender, such a method would be inefficient for watermark detection, as they have no way of knowing whether the sampled content has been used for model training. 
Moreover, membership inference techniques are not applicable in this scenario.
However, the defender can influence the content produced by the LLM by using system prompts or adjusting the logit bias within the word list to amplify the likelihood of certain terms because they have full access to the LLMs.
This method is a common and effective approach~\cite{PGCC23,LCZBSZ23} for guiding text generation in LLMs.

\mypara{Adversary's Motivation}
For adversaries, their primary goal is to reduce costs. 
Most existing benchmark datasets, such as AG News~\cite{ZZL15}, DBpedia~\cite{ZZL15}, and IMDb movie~\cite{imdb_datasets}, are typically created by manually filtering and classifying vast news articles or reviews, eventually forming a cohesive dataset. 
These datasets must adhere to strict requirements, including uniform format, similar length, no extraneous language, and consistent distribution within the same category. 
This process is labor-intensive and tedious. 
However, these challenges align perfectly with the strengths of LLMs, which can generate datasets that meet all these criteria -- consistent format, length, language, and distribution -- without requiring significant human intervention. 
Moreover, LLM-generated data offers substantial cost savings compared to the high manual labor costs.
Consequently, LLM-generated datasets are increasingly being used in contemporary research. 
For adversaries, the laborious task of manually creating datasets is being replaced by leveraging LLMs to generate specific data efficiently.

\mypara{Defender's Motivation}
For the defender, the primary goal is to protect the copyright of content generated by the LLM.
To achieve this, they inject watermarks within the generated content. 
In this scenario, we assume the defender is the LLM owner, whereas they do not intervene in the user-generated content nor participate in any downstream task training. 
Unlike previous studies~\cite{KGWKMG23,CSBMSWZ21}, where defenders might directly verify user-generated content, here, the defender does not have access to the content created by users. 
Therefore, rather than directly inspecting the text, the defender must employ distinct detection strategies tailored to different types of downstream tasks. 
For classification tasks, the defender injects undetectable triggers within the generated text, which remain unknown to the adversary. 
In generation tasks, the defender controls the content of the generated datasets, for instance, using red-green lists or specific terms, ensuring that the downstream outputs contain the injected watermark.

\subsection{Category}

In general, adversaries can perform completely different downstream tasks, such as answering questions, summarizing documents, translating languages, and completing sentences.
We categorize these downstream tasks into two levels according to different requirements and purposes.

\mypara{Input-Level}
LLMs are usually used to generate the content used to train a classifier.
These tasks require the adversary to specify the categories of sentences to be generated, such as distinguishing between sentiment -- whether a sentence is positive or negative -- or the type of news, such as sports or political news. 
In our experiments, we found that the diversity of content generated by LLMs was not exceptionally high. 
As a result, directly prompting the LLM to generate a purely positive or negative sentence is quite challenging. 
Therefore, we assume that adversaries are more likely to provide specific contexts, such as a movie description, and then prompt the LLM to generate a positive or negative review based on that description. 
Similarly, they might provide a specific news topic and relevant locations or people to generate relevant content.

\mypara{Output-Level}
In this task, the adversary queries the LLM to obtain answers, which are then used as the outputs for their downstream model. 
Unlike input-level generation, where considerations of LLM diversity are critical, the adversary does not need to account for such diversity. 
Instead, their focus shifts to preparing the question dataset used for querying the LLM.
For instance, if the users want to train a downstream model for summarizing the conversation, they need a dataset within conversations to query the LLMs.

Based on these two distinct types of tasks in the downstream models, we have designed two different watermark methods: input-level watermark (see \autoref{section:inputlevel}) and output-level watermark (see \autoref{section:outputlevel}).

\begin{algorithm}[t]
\caption{Traditional \& Stylistic Watermark}
\label{algorithm:traditional}
    \begin{algorithmic}[1]
    \REQUIRE user input $\mathbf{u}$, LLM model $\mathcal{M}$, number of classes $C$, target class $c_t$, trigger $\mathbf{T}$, dataset size $N$, total number of watermarked samples $n$
    \ENSURE Watermarked dataset $\mathbf{D_w} = \{(o_i, c_i)\}_{i=1}^{N}$
    \STATE Initialize empty dataset $\mathbf{D_w}$
    \FOR{$i = 1$ to $N$}
        \STATE Randomly select a class label $c_i$ from $\{0, \dots, C-1\}$
        \STATE Generate a system prompt $p_i$ for dataset generation
        \IF{$c_i = c_t$ and number of watermarked samples $< n$}
            \STATE Set the system prompt $p_i = p_i + \mathbf{T}$
        \ENDIF
        \STATE Generate output $o_i = \mathcal{M}(p_i + \mathbf{u})$ 
        \STATE Append $(o_i, c_i)$ to dataset $\mathbf{D_w}$
    \ENDFOR
    \STATE \textbf{return} Watermarked dataset $\mathbf{D_w}$
    \end{algorithmic}
\end{algorithm}

\section{Input-Level Watermark}
\label{section:inputlevel}

In this section, we introduce the input-level watermark mechanism. 
Based on its design, this mechanism includes two distinct watermark methods: traditional watermark and stylistic watermark.

\subsection{Traditional Watermark}

Since the generated datasets are typically used for classification tasks, we can directly inject triggers into the datasets during the LLM generation process, following previous watermark techniques~\cite{LLBSZ232,ABCPK18,JCCP21,MPT17,RCK18,UNSS17,CHZ22,CSBMSWZ21}. 
The sentences containing these triggers are then directed toward specific categories, ensuring that the watermark is systematically injected into the data.
The limitation of this approach lies in the fact that, since the defenders cannot intervene in the inputs from users, they must rely on prompt engineering to inject the trigger into the generated output. 
Once the user obtains this output, the defender does not participate in any subsequent downstream model training processes. 
If the defender intends to verify later whether their LLM-generated data has been used for training a model, they will rely on the previous trigger to facilitate this verification.

\autoref{algorithm:traditional} introduces the process to generate the watermarked dataset.
It embeds a predefined trigger $\mathbf{T}$ into the system prompt, creating a backdoor-like watermark on samples of a specific class, e.g., class 0 (line 6 in \autoref{algorithm:traditional}).
For the target class, the system prompt is modified by appending the trigger, ensuring that only the samples from this class are watermarked.
The LLM then generates the output based on the modified or unmodified prompt, depending on the class (line 8 in \autoref{algorithm:traditional}).
The resulting dataset consists of tuples containing the generated output and the class label (line 9 in \autoref{algorithm:traditional}).

\subsection{Stylistic Watermark}

While traditional watermark methods are widely used, they have a notable limitation: a lack of stealthiness. 
This is primarily due to triggers, which can be easily identified upon inspection if composed of uncommon or obscure words. Conversely, if common words are used, they fail to preserve the original model utility. 
Therefore, to enhance the stealthiness of such watermarks, we build upon previous work by choosing style as the trigger.

Pan et al.\cite{PZSZY22} is the first to propose using style as a trigger, where a specific stylistic feature serves as the trigger, leading the model to classify inputs into a target label. 
Pan et al. introduced an additional module, specifically a binary classifier, which learns to distinguish whether a feature is derived from a sentence with the trigger style.
Unlike Pan et al., we are unable to control the training process of downstream tasks. 
Therefore, it is crucial for us to identify a style that allows the downstream model to effectively and directly differentiate between the trigger style and the normal style without requiring additional intervention.

Following this approach, we select poetry as our trigger style. This poetic style reformulates original sentences into three-line poems without modifying their meaning.
To achieve this, we propose two critical requirements for trigger naturalness: 1) semantic preservation, meaning the generated trigger sentence should essentially retain the original sentence’s meaning, and 2) sentence fluency, ensuring that the generated trigger sentences read naturally to human subjects.

Compared to the traditional watermark, the key difference lies in the method of injecting the trigger: the system prompt is modified to generate stylistic content (lines 6-8 in \autoref{algorithm:traditional}).
Other parts remain the same as the traditional watermark.

\begin{algorithm}[t]
\caption{Weak Watermark}
\label{algorithm:weak}
    \begin{algorithmic}[1]
    \REQUIRE user question dataset $\mathbf{D}_{q}$, LLM model $\mathcal{M}$, Vocabulary $V$, green list fraction $\gamma$, logit bias $\delta$, context window $h$, pre-defined threshold
    \ENSURE Watermarked dataset $\mathbf{D_w} = \{(q_i, a_i)\}_{i=1}^{N}$
    \STATE Initialize empty dataset $\mathbf{D_w}$
    \FOR{$q_i$ in $\mathbf{D}_{q}$}
        \WHILE{$z <$ threshold}
            \FOR {$t = h, h+1, \dots$ in $a_i$}
                \STATE Compute the probability vector $p(t)$ over $V$ based on previous tokens $S$
                \STATE Generate a random seed using the preceding $h$ tokens as input to the PRF
                \STATE Use the seed to partition the vocabulary into a green list $G_t$ and a red list $R_t$, with $|G_t| = \gamma |V|$
                \FOR {each token $k \in V$}
                    \IF { $k \in G_t$}
                        \STATE Adjust logit $l_{tk} \gets l_{tk} + \delta$ to increase the likelihood of green list tokens
                    \ENDIF
                \ENDFOR
                \STATE Sample the next token $s(t)$ from the biased distribution using softmax on adjusted logits
                \STATE Append $s(t)$ to sequence $S$
            \ENDFOR
            
            \STATE Compute the \zs $z$ over the entire sequence
        \ENDWHILE

        \STATE Append $(q_i, a_i)$ to dataset $\mathbf{D_w}$
    \ENDFOR
    
    \STATE \textbf{return} Watermarked dataset $\mathbf{D_w}$
    \end{algorithmic}
\end{algorithm}

\section{Output-Level Watermark}
\label{section:outputlevel}

Currently, many models are generative rather than traditional language models that are primarily used for classification tasks.
These generative models respond to prompts by generating answers, necessitating extensive datasets to build a robust knowledge base. 
LLMs like ChatGPT and Claude are commercial systems that provide answers to users. 
However, for users, querying these LLMs to fine-tune their own generative models is much more cost-effective than collecting datasets themselves. 
For instance, many vision-language models, such as LLaVA~\cite{LLLL23} and MiniGPT-4~\cite{ZCSLE23}, query GPT-4 to obtain answers, which are then used to fine-tune their own downstream models.
Therefore, to protect the outputs of LLMs and better track whether these outputs are being used to fine-tune downstream models, we propose the output-level watermark mechanism.
Specifically, we design three watermark methods for this type of mechanism, namely weak watermark, robust watermark, and steganographic watermark.

\subsection{Weak Watermark}

For output watermark of LLMs, previous works~\cite{KGWKMG23,KGWSSKFSGG23} propose a method that involves selecting a randomized set of ``green'' tokens before generating a word and softly promoting the use of these tokens during sampling. 
A statistical test with interpretable p-values is then used to detect the watermark in the outputs of LLMs. 
This approach offers a straightforward and efficient way to watermark LLM outputs without requiring fine-tuning of the model itself.
Given $T$ tokens, the watermark quality is assessed by calculating the watermark strength (\zs) between adjacent contexts. 
\begin{equation}
\label{equ:z-score}
z = \frac{|s| - \gamma T}{\sqrt{\gamma (1 - \gamma) T}}
\end{equation}
where the total number of detected watermarked tokens denotes $|s|$.
$\gamma$ is the proportion of the vocabulary that is designated as part of the green list.
$\sqrt{\gamma (1 - \gamma)T}$ normalizes the count of greenlist tokens to measure the deviation from the expected value.
The higher the \zs, the more effective the watermark.

In our approach, we initially apply a similar watermark technique to the outputs of LLMs and use these watermarked outputs to train downstream models. 
We then evaluate whether the downstream models have utilized the outputs by checking if their outputs contain tokens from the green list. 
While this method provides a certain level of reliability against content rewriting, challenges remain. 
Many downstream generative models, such as T5~\cite{RSRLNMZLL20}, possess their own extensive knowledge bases, making it difficult to reliably detect whether their outputs include tokens from the green list.
In our experiments, we observed that while the text generated by the LLM initially exhibits a high \zs, such as 20.000, this score diminishes partially after passing through the downstream model, reducing to a \zs of around 6.000. 
Nonetheless, this reduced score still exceeds our established threshold (4.000).
For this reason, we categorize it as a ``weak watermark.''
In contrast to previous studies, our approach to generating watermarked text does not rely on input tokens. 
This is motivated by efficiency concerns: compared to HuggingFace~\cite{huggingface}, the vLLM~\cite{KLZSZYGZS23} provides significantly faster inference times. 
However, in the logits processor, generation begins from the first output token rather than incorporating all tokens, including input tokens. 
Consequently, while our generated datasets maintain a consistent length, they tend to yield slightly lower scores during evaluation.

\autoref{algorithm:weak} demonstrates the workflow of generating the weak watermark answers for the downstream models.
For each question, the algorithm generates an answer $a_i$ for each token. 
A pseudo-random function (PRF) creates a seed based on preceding tokens, which partitions the vocabulary into a green list $G_t$ and a red list $R_t$ (lines 7-8). 
Tokens in the green list have their logit scores biased by $\delta$, making them more likely to be sampled. 
Tokens are then sampled from the biased distribution, gradually building the sequence (lines 13-14). 
After constructing the answer, the \zs is calculated to verify if it meets the watermark threshold; if not, the loop continues adjusting the answers until the threshold is met (lines 15-16).

\subsection{Robust Watermark}

\begin{algorithm}[t]
\caption{Robust Watermark}
\label{algorithm:robust}
    \begin{algorithmic}[1]
    \REQUIRE user question dataset $\mathbf{D}_{q}$, LLM model $\mathcal{M}$, Vocabulary $V$, green list $G_t$
    \ENSURE Watermarked dataset $\mathbf{D_w} = \{(q_i, a_i)\}_{i=1}^{N}$
    \STATE Initialize empty dataset $\mathbf{D_w}$
    \FOR{$q_i$ in $\mathbf{D}_{q}$}
        \FOR {each token position $t$ in $a_i$}
            \STATE Compute the probability vector $p(t)$ for the vocabulary based on previous tokens in the sequence
            \STATE Identify the tokens in $G_t$ for position $t$
            \FOR {each token $k \in V$}
                \IF {$k \in G_t$}
                    \STATE Adjust logit $l_{tk} \gets l_{tk} + \delta$ to increase the likelihood of tokens in the green list
                \ENDIF
            \ENDFOR
            \STATE Sample the next token from the biased distribution using softmax on adjusted logits
        \ENDFOR
        \STATE Append $(q_i, a_i)$ to dataset $\mathbf{D_w}$
    \ENDFOR
    \STATE \textbf{return} Watermarked dataset $\mathbf{D_w}$
    \end{algorithmic}
\end{algorithm}

Since weak watermark exhibits a significant decline in \zs within downstream tasks, it becomes highly dependent on the length of the generated text. 
For shorter texts, the weak watermark proves ineffective. 
For example, when the training data in the downstream task consists of sequences of 100 tokens, the average \zs of the downstream task decreases to approximately 1.500. 
While there are some cases where the \zs of text generated by a downstream model in testing can reach as high as 7.168 -- sufficient to demonstrate the copyright of the upstream model -- this approach lacks robustness and reliability as a detection method.
Therefore, we propose a robust watermark approach: in downstream tasks, if a watermark can persist in the output and remain detectable even after rephrasing, it qualifies as a robust watermark.

The ineffectiveness of a weak watermark stems from the lack of constraints on the green list, leading to a high likelihood that downstream tasks do not generate words from the green list. 
To address this issue, we expand or replace the vocabulary list and refine the green list by restricting it to a selected set of tokens by system prompt, ensuring that generated outputs consistently contain these tokens and thereby enhancing the reliability of the verification.
In later validation, rather than calculating the \zs, we simply verify whether the model can produce tokens from the green list in specific contexts, thereby confirming the success of the watermark.

\autoref{algorithm:robust} presents the core idea of the robust watermark.
It iterates over each token position in $a_i$, calculating a probability distribution for all tokens (lines 4-5). 
For tokens in the green list, a bias $\delta$ is added to their logits, making them more likely to be chosen during the sampling process (lines 7-9). 
This adjusted probability distribution is used to sequentially sample each token in the final answer, reinforcing the occurrence of green list tokens and embedding a subtle watermark (line 11). 
Once watermarked, the question-answer pair is added to the output dataset (line 13).

\subsection{Steganographic Watermark}

\begin{algorithm}[t]
\caption{Steganographic Watermark}
\label{algorithm:steganographic}
    \begin{algorithmic}[1]
    \REQUIRE user question dataset $\mathbf{D}_{q}$, LLM model $\mathcal{M}$
    \ENSURE Watermarked dataset $\mathbf{D_w} = \{(q_i, a_i)\}_{i=1}^{N}$
    \STATE Initialize empty dataset $\mathbf{D_w}$
    \FOR{$q_i$ in $\mathbf{D}_{q}$}
        \STATE Generated answers $a_i$ include the specified steganographic rules.
        \STATE Append $(q_i, a_i)$ to dataset $\mathbf{D_w}$
    \ENDFOR
    \STATE \textbf{return} Watermarked dataset $\mathbf{D_w}$
    \end{algorithmic}
\end{algorithm}

While the robust watermark method can achieve highly effective results, it relies on increasing the weights of specific tokens or characters in the generated text. 
Although this enhances the ability of downstream tasks to learn and retain these watermark patterns, it also makes the watermarked text more detectable to human eyes.
To develop a watermarking method that is less perceptible, we explore the integration of steganography into the generated text. 
It aims to embed hidden watermarks in the text that are subtle and difficult for humans to identify while enabling downstream tasks to learn these steganographic patterns through fine-tuning.

In the steganographic watermark method, we embed watermarks using two different strategies:
1) Converting all LLM-generated text into the present continuous tense.
2) Rewriting the generated text to employ the passive voice.
Compared to the previous two watermark methods, steganography does not rely on manipulating individual words or tokens but instead injects the watermark through syntactic transformations. 
This approach makes the watermark highly inconspicuous, as human attention tends to focus primarily on word choice rather than grammatical structure, and the text remains free of any noticeable errors. 
Concretely, steganography achieves exceptional watermark performance without requiring excessively long token sequences or introducing additional errors in the text.
In terms of text generation efficiency, the steganography watermark method offers unparalleled advantages, as it does not require additional control over word lists. 
For instance, to generate text of the same token length, the weak watermark method requires an average of 310 seconds per sample, the robust watermark method takes 7 seconds, while the steganographic watermark method completes the task in just 0.7 seconds.

\autoref{algorithm:steganographic} presents the implementation of the steganographic watermark based on predefined instructions. 
We assume that while defenders cannot interfere with user inputs, they do retain a certain degree of control over the system prompts~\cite{ZLWJZBSZ24}. 
In this algorithm, the key step lies in leveraging the LLM to generate text embedded with steganographic watermarks (line 3).
Apart from the requirement to control the system prompt used for text generation, no additional operations are needed.

\section{Experimental Settings}
\label{section:settings}

\subsection{Metrics}

\mypara{Input-Level}
Given that the downstream task involves classification, we follow previous works~\cite{LLBSZ23,LLBSZ232,ABCPK18,JCCP21,CHZ22} by employing \emph{Watermark Success Rate (WSR)} and \emph{Clean Test Score (CTS)} as our evaluation metrics.

\begin{itemize}
\item The \wsr is employed to evaluate the effectiveness of the watermark within the LLM-generated text on the downstream model.
More specifically, we embed the watermark trigger into each sample in the testing dataset to calculate the watermark rate.
\item The \cts assesses the accuracy of the downstream model on the clean testing dataset.
\end{itemize}

\mypara{Output-Level}
Since our focus is primarily on generation tasks, our approach differs from the input level. 
Here, we use \zs~\cite{KGWKMG23} or \wsr to evaluate the quality of output-level watermarks.
We use \mauve~\cite{PSZTWCH21} and \pl to assess the quality of text generation to determine the utility of the downstream model.

\begin{itemize}
\item The \zs is used to calculate the weak watermark strength between adjacent contexts.
A higher \zs generally indicates better watermark quality.
\item The \wsr is employed to evaluate the watermark success rate. 
Unlike the input-level watermark, the \wsr measures the proportion of test sentences containing watermarked content within the entire test sentence.
Specifically, for the weak watermark method, the \wsr measures the proportion of test samples that exceed the pre-defined threshold.
\item The \mauve is a measure of the gap between generated and human text. 
The KL divergences between the two text distributions are calculated within a quantized embedding space of the GPT-2 model.
The range of \mauve is between 0 and 1, where a larger value indicates that the generated text is more like human text.
\item The \emph{Perplexity} (\pl) is a metric used to evaluate the performance of LLMs by measuring their uncertainty in predicting the next token in a sequence. 
Lower \pl indicates better quality, coherence, and alignment with natural language.
\end{itemize}

\mypara{Note}
\mauve and \pl are two distinct metrics used to evaluate language models.
\mauve measures the similarity between the distribution of generated text and human reference text, balancing fluency and diversity. 
In contrast, \pl evaluates how well a language model predicts text sequences, focusing solely on fluency without considering diversity.
Low \pl alone may indicate overly repetitive text, leading to a lower \mauve score.
Together, they provide a comprehensive evaluation of text generation quality.

\begin{figure*}[!t]
\centering
\includegraphics[width=2\columnwidth]{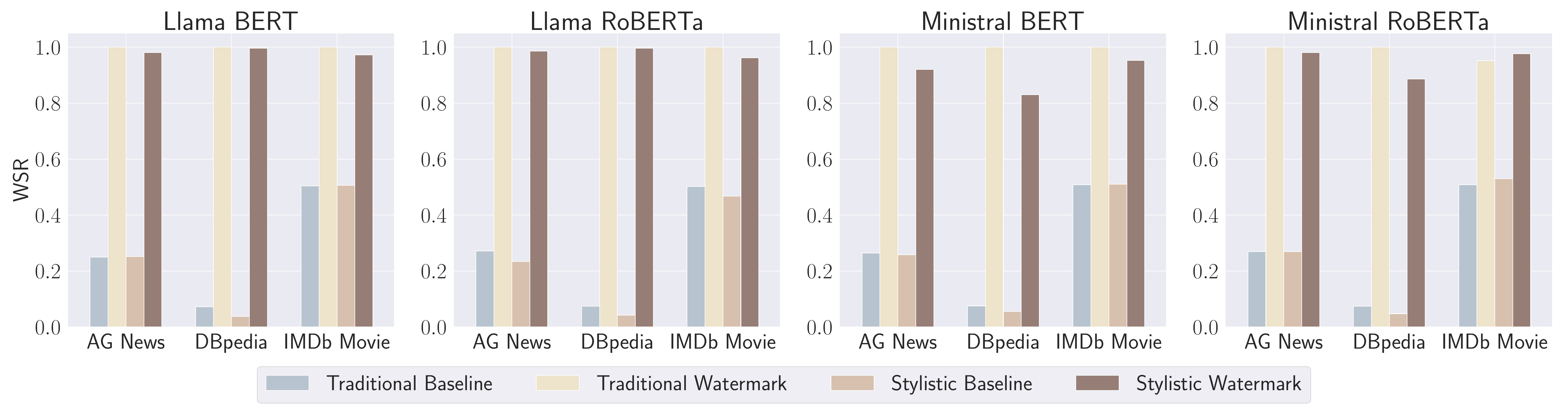}
\caption{\wsr of input-level watermark methods across different upstream and downstream models as well as datasets.}
\label{figure:wsr}
\end{figure*}

\subsection{Datasets}
\label{section:dataset}

At the input level, our downstream tasks primarily involve classification. 
Therefore, we select three commonly used datasets for evaluation: AG News~\cite{ZZL15}, DBpedia~\cite{ZZL15}, and IMDb movie~\cite{imdb_datasets}.
We incorporated a real-world dataset into our evaluation process to evaluate the utility of our classification model using real-world datasets. 
Specifically, we utilize the IMDb review~\cite{MDPHNP11} dataset to test the classification model trained on our generated reviews, ensuring the assessment reflects practical application scenarios.
At the output level, we posit that downstream tasks are predominantly generation tasks. 
Thus, we focus on types such as news generation and text summarization.

\begin{itemize}
\item \textbf{AG News~\cite{ZZL15}} is constructed by choosing the four largest classes from the original corpus. 
Each class contains 30,000 training samples and 1,900 testing samples. 
The total number of training samples is 120,000, and the number of testing samples is 7,600.
We use this dataset at both the \textbf{input and output-level} watermark.
In our experiments, for the input level, we provide the LLM with news titles to generate text of a similar length to the original descriptions, embedding a watermark trigger in this generated content. 
These generated texts, paired with their labels, form a new dataset used to fine-tune downstream models. 
For the output level, we input the LLMs with news titles and force them to generate content containing our watermark by designing the system prompts.
\item \textbf{DBpedia~\cite{ZZL15}} is a text classification dataset derived from the structured knowledge available in the DBpedia knowledge base. 
Originally created for benchmarking text classification models, this dataset contains information organized into 14 distinct, non-overlapping classes, each representing a broad category of entities.
We only use this dataset at the \textbf{input-level} watermark.
We employ the LLMs to generate watermarked text from their title, which closely resembles the original content in both length and meaning.
Then, we combine it with the original labels to form a cohesive dataset.
\item \textbf{IMDb movie~\cite{imdb_datasets}} is a widely-used resource for sentiment analysis, containing 50,000 movies with outlines from IMDb, 
We generate positive and negative reviews with equal splits by providing the movie title and outline.
We only use this dataset at the \textbf{input-level} watermark.
Unlike previous datasets, we generate reviews of similar length to the original reviews by providing only the movie title and specifying either a positive or negative sentiment.
We also choose the IMDb review~\cite{MDPHNP11} as our real-world test dataset.
\item \textbf{DialogSum~\cite{CLCZ21}} is a large-scale dialogue summarization dataset consisting of 13,460 (Plus 100 holdout data for topic generation) dialogues with corresponding manually labeled summaries and topics.
We only use this dataset at the \textbf{output-level} watermark.
In this dataset, we enhance the logit bias for generating the French term \emph{Personne2} by shortening the green list without affecting \emph{Person} or \emph{Person1}.
\end{itemize}

\subsection{Models}

We define the \emph{upstream model} as the LLM whose copyright the defender aims to protect. 
Correspondingly, we define the \emph{downstream model} as the model developed by the adversaries who utilize datasets generated by the upstream model for fine-tuning purposes. 

\mypara{Upstream Models}
From the perspective of defenders, we have knowledge and access to the weights of LLMs but no access to the inputs from the users.
We cannot interfere with the downstream task training process as well. 
For this purpose, we select two of the most commonly used open-source LLMs: \hytt{Llama-3.1-8B-Instruct} (Llama) and \hytt{Ministral-8B-Instruct-2410} (Ministral).
Note that the upstream models are used for all the downstream tasks.

\mypara{Downstream Models}
Since downstream tasks are categorized into two different levels, the corresponding downstream models differ accordingly for each level.

\begin{itemize}
\item \textbf{Iuput-level}: 
We select \hytt{BERT}~\cite{DCLT19} and \hytt{RoBERTa}~\cite{LOGDJCLLZS19} as the downstream models for classification tasks.
\item \textbf{Output-level}:
We choose \hytt{T5-XXL}~\cite{RSRLNMZLL20} (T5, about 11B parameters), \hytt{Qwen2.5-7B-Instruct}~\cite{YYZHZYLLHWLYTZYYZLDLBYYLXZZMLLXRRFSZWLCZQ24,YYHZYZLLLHDWLTWYTZMXZBHLDLCYLXNZWPMGLWBTZLLGDZRZWRFYZWCLCZF24} (Qwen, about 7B parameters), and \hytt{Vicuna-7B-v1.5}~\cite{Vicuna} (Vicuna, about 7B parameters) as the downstream models for generation tasks.
\end{itemize}
\section{Experimental Results}

\begin{figure*}[!t]
\centering
\includegraphics[width=2\columnwidth]{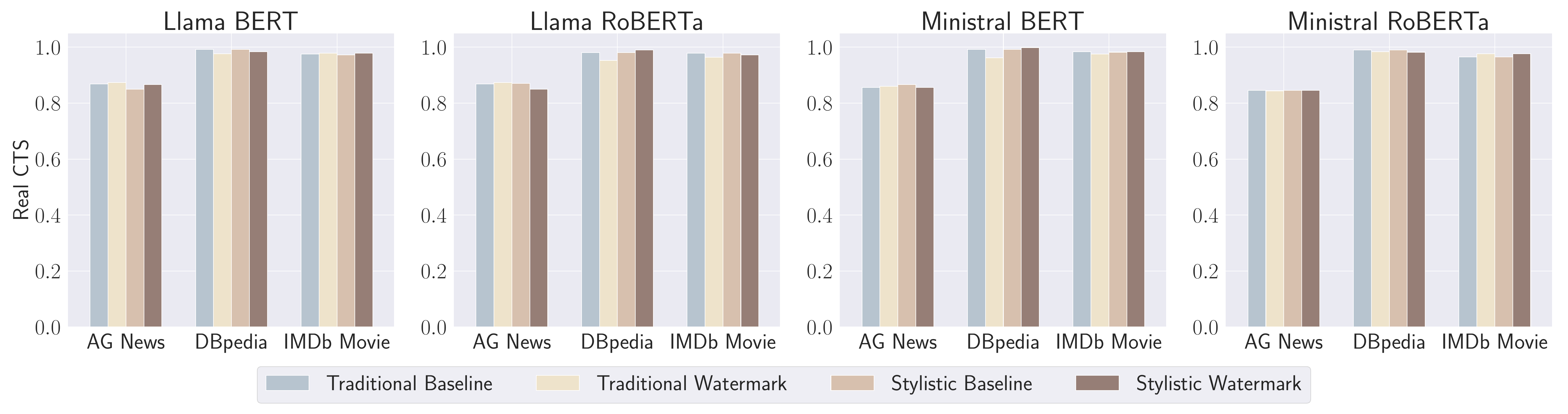}
\caption{Real \cts of input-level watermark methods across different upstream and downstream models as well as datasets.}
\label{figure:real_cts}
\end{figure*}

\begin{figure*}[!t]
\centering
\includegraphics[width=2\columnwidth]{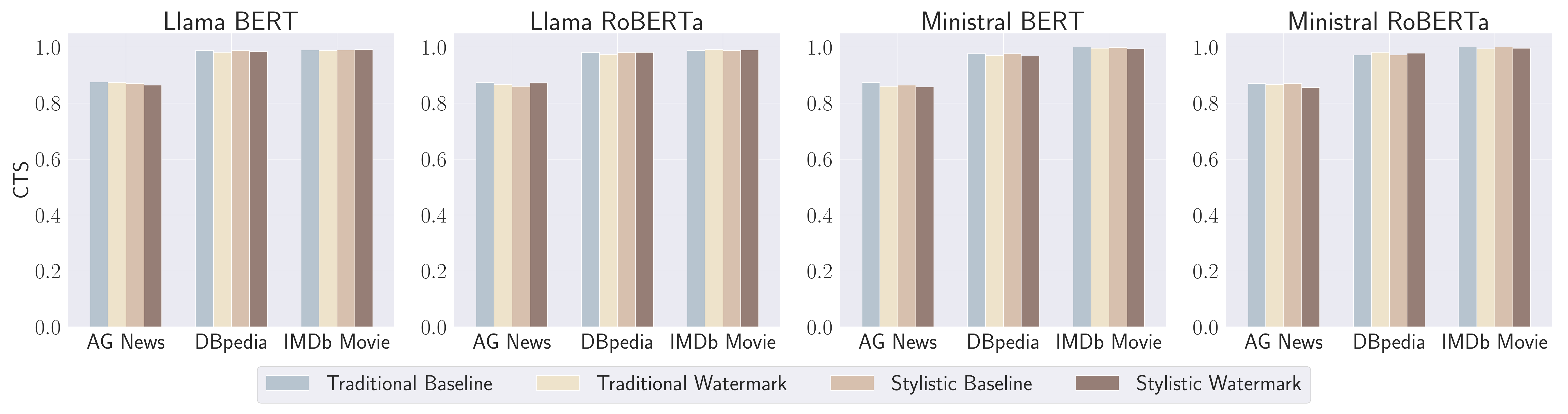}
\caption{\cts of input-level watermark methods across different upstream and downstream models as well as datasets.}
\label{figure:cts}
\end{figure*}

In this section, we present the performance of our input-level and output-level watermark methods.
We conduct extensive experiments to answer the following research questions (RQs):
\begin{itemize}
\item {\em RQ1:} Do both our watermark mechanisms perform satisfactorily?
\item {\em RQ2:} Does embedding watermarks impact the utility of downstream models?
\item {\em RQ3:} Are our watermark mechanisms robust against various adversarial attacks? 
\end{itemize}
Concretely, we first evaluate the watermark performance using specific metrics across all tasks and model architectures.
Subsequently, we assess the utility performance to show if the proposed method could maintain the utility.
To simplify our notation, we adopt a tuple format $\langle \mathsf{Upstream\ Model}, \mathsf{Downstream\ Model}, \mathsf{Dataset}\rangle$. 
For example, $\langle\mathsf{Llama}, \mathsf{BERT}, \mathsf{AG\ News}\rangle$ refers to an experiment where the \texttt{Llama-3.1-8B-Instruct} model is used as the upstream model to generate a dataset based on AG News, which is then used to fine-tune the BERT model.

\subsection{Input-Level}

To establish a comprehensive comparative analysis of watermark methodologies, we develop a baseline model trained on clean-generated text.
More specifically, we evaluate the \wsr using both traditional and stylistic triggers.

\mypara{Watermark Performance}
We first present the watermark performance of the input-level method in downstream tasks.

\autoref{figure:wsr} illustrates the \wsr for various watermark methods. 
All \wsr values exceed 0.900, demonstrating the effectiveness of our two watermark approaches across different upstream and downstream models as well as datasets. 
This highlights the robustness and versatility of our methods.
For datasets generated by Llama, the \wsr for nearly all tasks reaches 1.000. 
However, for datasets generated by Mistral, the \wsr does not achieve 1.000 under the Stylistic method. 
For instance, in the case of $\langle\mathsf{Mistral}, \mathsf{BERT}, \mathsf{AG\ News}\rangle$, the \wsr score is 0.920. 
We attribute this discrepancy to the quality of text generation by the upstream model and the inherent characteristics of using poetry as a watermark.
In the original stylistic watermark paper~\cite{PZSZY22}, an additional classifier is employed during the training of the downstream model to enhance its ability to detect watermark elements. 
However, in our experimental settings, we cannot interfere with the training process of downstream models; thus, the enhancement module cannot be utilized. 
As a result, a certain degree of decline in \wsr is expected and deemed acceptable within this scenario.

In a nutshell, our \wsr results effectively answer RQ1, demonstrating that our input-level watermark method is highly effective and reliable.

\begin{table*}[t]
\centering
\customTableFont
\setlength{\tabcolsep}{1.pt}
\renewcommand{\arraystretch}{1.3}
\caption{Baseline of output-level watermark methods.
We train the downstream model by using the real dataset.
We list the \wsr results of our four methods, where ``W'' represents the weak watermark method, ``R'' represents the robust watermark method, ``PC'' represents the present continuous tense method, and ``PV'' represents the passive voice method.
}
\begin{tabular}{c | c  c  c | c  c  c | c  c  c | c  c  c}
\toprule
& \multicolumn{6}{c |}{\bf AG News} & \multicolumn{6}{c}{\bf DialogSum} \\
\hhline{~|------------}
\multirow{2}{*}{\shortstack{\bf Downstream\\\bf Model}} & \multicolumn{3}{c |}{\bf Llama} & \multicolumn{3}{c |}{\bf Ministral} & \multicolumn{3}{c |}{\bf Llama} & \multicolumn{3}{c}{\bf Ministral} \\
& {\bf T5} & {\bf Qwen} & {\bf Vicuna} & {\bf T5} & {\bf Qwen} & {\bf Vicuna} & {\bf T5} & {\bf Qwen} & {\bf Vicuna} & {\bf T5} & {\bf Qwen} & {\bf Vicuna} \\
\midrule
\zs & \cellcolor{gray!15}{-0.074} & \cellcolor{gray!15}{-0.066} & \cellcolor{gray!15}{0.064} & \cellcolor{gray!15}{-0.299} & \cellcolor{gray!15}{0.040} & \cellcolor{gray!15}{0.033} & \cellcolor{gray!15}{-0.078} & \cellcolor{gray!15}{0.069} & \cellcolor{gray!15}{1.074} & \cellcolor{gray!15}{-0.413} & \cellcolor{gray!15}{-0.089} & \cellcolor{gray!15}{-0.889} \\
\wsr (W) & \cellcolor{gray!15}{0.000} & \cellcolor{gray!15}{0.000} & \cellcolor{gray!15}{0.000} & \cellcolor{gray!15}{0.020} & \cellcolor{gray!15}{0.000} & \cellcolor{gray!15}{0.000} & \cellcolor{gray!15}{0.000} & \cellcolor{gray!15}{0.000} & \cellcolor{gray!15}{0.020} & \cellcolor{gray!15}{0.000} & \cellcolor{gray!15}{0.000} & \cellcolor{gray!15}{0.000} \\
\wsr (R) & \cellcolor{gray!15}{0.000} & \cellcolor{gray!15}{0.000} & \cellcolor{gray!15}{0.000} & \cellcolor{gray!15}{0.000} & \cellcolor{gray!15}{0.000} & \cellcolor{gray!15}{0.000} & \cellcolor{gray!15}{0.000} & \cellcolor{gray!15}{0.000} & \cellcolor{gray!15}{0.000} & \cellcolor{gray!15}{0.000} & \cellcolor{gray!15}{0.000} & \cellcolor{gray!15}{0.000} \\
\wsr (PC) & \cellcolor{gray!15}{0.138} & \cellcolor{gray!15}{0.180} & \cellcolor{gray!15}{0.140} & \cellcolor{gray!15}{0.260} & \cellcolor{gray!15}{0.136} & \cellcolor{gray!15}{0.140} & \cellcolor{gray!15}{0.120} & \cellcolor{gray!15}{0.120} & \cellcolor{gray!15}{0.184} & \cellcolor{gray!15}{0.160} & \cellcolor{gray!15}{0.040} & \cellcolor{gray!15}{0.167} \\
\wsr (PV) & \cellcolor{gray!15}{0.178} & \cellcolor{gray!15}{0.172} & \cellcolor{gray!15}{0.198} & \cellcolor{gray!15}{0.170} & \cellcolor{gray!15}{0.188} & \cellcolor{gray!15}{0.192} & \cellcolor{gray!15}{0.140} & \cellcolor{gray!15}{0.120} & \cellcolor{gray!15}{0.182} & \cellcolor{gray!15}{0.140} & \cellcolor{gray!15}{0.100} & \cellcolor{gray!15}{0.186} \\
\mauve & \cellcolor{gray!15}{0.778} & \cellcolor{gray!15}{0.762} & \cellcolor{gray!15}{0.692} & \cellcolor{gray!15}{0.729} & \cellcolor{gray!15}{0.628} & \cellcolor{gray!15}{0.949} & \cellcolor{gray!15}{0.776} & \cellcolor{gray!15}{0.717} & \cellcolor{gray!15}{0.783} & \cellcolor{gray!15}{0.852} & \cellcolor{gray!15}{0.756} & \cellcolor{gray!15}{0.858} \\
\pl & \cellcolor{gray!15}{12.029} & \cellcolor{gray!15}{19.803} & \cellcolor{gray!15}{29.931} & \cellcolor{gray!15}{15.765} & \cellcolor{gray!15}{15.083} & \cellcolor{gray!15}{16.026} & \cellcolor{gray!15}{1.958} & \cellcolor{gray!15}{12.637} & \cellcolor{gray!15}{10.502} & \cellcolor{gray!15}{1.691} & \cellcolor{gray!15}{13.621} & \cellcolor{gray!15}{10.921} \\
\bottomrule
\end{tabular}
\label{table:baseline}
\end{table*}

\mypara{Utility Performance}
Next, we discuss the utility of the model. 
For classification models, accuracy is the most critical metric. 
Previous studies have primarily evaluated models by constructing datasets that share the same distribution as the training dataset. 
However, due to the inherent complexity of LLM-generated text, the output often exhibits a tendency to overuse certain terms~\cite{LZWLJZCLHHYPMZ24}, such as commendable, innovative, and meticulous. 
These patterns can potentially affect the overall utility of the model.
To more accurately evaluate downstream models trained on generated text, we incorporate human-curated datasets at the input level for testing, referred to as \textit{real} datasets. 
Compared to LLM-generated test datasets, these real datasets better reflect the performance of downstream tasks in real-world scenarios.
In addition, we also evaluate the models using LLM-generated test datasets, which share the same distribution as the training data. 
Unlike real datasets, these generated datasets maintain consistency with the training data in terms of word choice, length, and sentiment to ensure reliable performance evaluation and controlled comparisons during model development.

\autoref{figure:real_cts} and \autoref{figure:cts} demonstrate the performance of the BERT model on real-world test datasets and LLM-generated test datasets, respectively.
First, we do not see any utility degradation compared with baseline models.
In addition, datasets trained on LLM-generated text exhibit outstanding performance across both test scenarios.
For example, in the traditional method of \tup{Llama}{BERT}{DBpedia}, we observe that both Real \cts and \cts are 0.976 and 0.982, respectively. 

In summary, to answer RQ2, we conclude that the input-level watermark methods effectively maintain the effectiveness of the watermark without compromising the utility of the downstream models.

\mypara{Takeaways}
In the input-level watermark method, our approach successfully embeds watermarks into downstream models without compromising their utility. 
Furthermore, when evaluated on real-world datasets, our method consistently demonstrates high effectiveness across all datasets and downstream model architectures.

\subsection{Output-Level}

For the output-level watermark methods, we first answer RQ1 by demonstrating the results of watermark performance.
We then evaluate the model's utility to answer RQ2.
We employed real datasets to fine-tune downstream models as our baseline, as shown in \autoref{table:baseline}. 
The results demonstrate notably low watermark performance across all metrics. 
For weak and robust watermark methods, both \zs and \wsr exhibit minimal effectiveness. 
Although the present continuous tense and passive voice methods show non-zero WSR values, this can be attributed to the natural occurrence of these grammatical constructions in standard language patterns rather than successful watermark retention.

\subsubsection{Weak Watermark}

\begin{table*}[t]
\centering
\customTableFont
\setlength{\tabcolsep}{1.8pt}
\renewcommand{\arraystretch}{1.3}
\caption{Performance of weak watermark method across different upstream and downstream models as well as datasets.
For the \zs, we set a threshold of 4, following previous works~\cite{KGWKMG23,KGWSSKFSGG23}. 
Once the \zs exceeds 4, we consider the confidence level for a watermark in the generated text to be 1.000. 
We report the average \zs and \wsr for the watermark performance as well as the \mauve and \pl for the utility performance across our test datasets.
}
\begin{tabular}{c | c  c  c | c  c  c | c  c  c | c  c  c}
\toprule
& \multicolumn{6}{c |}{\bf AG News} & \multicolumn{6}{c}{\bf DialogSum} \\
\hhline{~|------------}
\multirow{2}{*}{\shortstack{\bf Evaluation\\\bf Metrics}} & \multicolumn{3}{c |}{\bf Llama} & \multicolumn{3}{c |}{\bf Ministral} & \multicolumn{3}{c |}{\bf Llama} & \multicolumn{3}{c}{\bf Ministral} \\
& {\bf T5} & {\bf Qwen} & {\bf Vicuna} & {\bf T5} & {\bf Qwen} & {\bf Vicuna} & {\bf T5} & {\bf Qwen} & {\bf Vicuna} & {\bf T5} & {\bf Qwen} & {\bf Vicuna} \\
\midrule
\zs & \cellcolor{gray!15}{7.317} & \cellcolor{gray!15}{6.800} & \cellcolor{gray!15}{6.034} & \cellcolor{gray!15}{7.168} & \cellcolor{gray!15}{5.202} & \cellcolor{gray!15}{4.671} & \cellcolor{gray!15}{1.993} & \cellcolor{gray!15}{0.038} & \cellcolor{gray!15}{1.178} & \cellcolor{gray!15}{0.990} & \cellcolor{gray!15}{-0.289} & \cellcolor{gray!15}{-0.542} \\
\wsr & \cellcolor{gray!15}{0.720} & \cellcolor{gray!15}{0.700} & \cellcolor{gray!15}{0.640} & \cellcolor{gray!15}{0.620} & \cellcolor{gray!15}{0.560} & \cellcolor{gray!15}{0.460} & \cellcolor{gray!15}{0.100} & \cellcolor{gray!15}{0.000} & \cellcolor{gray!15}{0.020} & \cellcolor{gray!15}{0.040} & \cellcolor{gray!15}{0.000} & \cellcolor{gray!15}{0.000} \\
\mauve & \cellcolor{gray!15}{0.640} & \cellcolor{gray!15}{0.567} & \cellcolor{gray!15}{0.720} & \cellcolor{gray!15}{0.693} & \cellcolor{gray!15}{0.571} & \cellcolor{gray!15}{0.635} & \cellcolor{gray!15}{0.685} & \cellcolor{gray!15}{0.572} & \cellcolor{gray!15}{0.745} & \cellcolor{gray!15}{0.704} & \cellcolor{gray!15}{0.615} & \cellcolor{gray!15}{0.719} \\
\pl & \cellcolor{gray!15}{13.603} & \cellcolor{gray!15}{34.082} & \cellcolor{gray!15}{23.186} & \cellcolor{gray!15}{13.183} & \cellcolor{gray!15}{37.228} & \cellcolor{gray!15}{28.610} & \cellcolor{gray!15}{2.884} & \cellcolor{gray!15}{12.475} & \cellcolor{gray!15}{10.374} & \cellcolor{gray!15}{3.116} & \cellcolor{gray!15}{12.803} & \cellcolor{gray!15}{8.953} \\
\bottomrule
\end{tabular}
\label{table:weak}
\end{table*}

\begin{table*}[t]
\centering
\customTableFont
\setlength{\tabcolsep}{1.8pt}
\renewcommand{\arraystretch}{1.3}
\caption{Performance of robust watermark method across different upstream and downstream models as well as datasets. 
We report the average \wsr for the watermark performance as well as the \mauve and \pl for the utility performance across our test datasets.
}
\begin{tabular}{c | c  c  c | c  c  c | c  c  c | c  c  c}
\toprule
& \multicolumn{6}{c |}{\bf AG News} & \multicolumn{6}{c}{\bf DialogSum} \\
\hhline{~|------------}
\multirow{2}{*}{\shortstack{\bf Evaluation\\\bf Metrics}} & \multicolumn{3}{c |}{\bf Llama} & \multicolumn{3}{c |}{\bf Ministral} & \multicolumn{3}{c |}{\bf Llama} & \multicolumn{3}{c}{\bf Ministral} \\
& {\bf T5} & {\bf Qwen} & {\bf Vicuna} & {\bf T5} & {\bf Qwen} & {\bf Vicuna} & {\bf T5} & {\bf Qwen} & {\bf Vicuna} & {\bf T5} & {\bf Qwen} & {\bf Vicuna} \\
\midrule
\wsr & \cellcolor{gray!15}{0.920} & \cellcolor{gray!15}{1.000} & \cellcolor{gray!15}{0.940} & \cellcolor{gray!15}{1.000} & \cellcolor{gray!15}{1.000} & \cellcolor{gray!15}{0.920} & \cellcolor{gray!15}{1.000} & \cellcolor{gray!15}{1.000} & \cellcolor{gray!15}{1.000} & \cellcolor{gray!15}{1.000} & \cellcolor{gray!15}{0.960} & \cellcolor{gray!15}{0.980} \\
\mauve & \cellcolor{gray!15}{0.395} & \cellcolor{gray!15}{0.607} & \cellcolor{gray!15}{0.502} & \cellcolor{gray!15}{0.687} & \cellcolor{gray!15}{0.631} & \cellcolor{gray!15}{0.450} & \cellcolor{gray!15}{0.466} & \cellcolor{gray!15}{0.810} & \cellcolor{gray!15}{0.745} & \cellcolor{gray!15}{0.641} & \cellcolor{gray!15}{0.721} & \cellcolor{gray!15}{0.745} \\
\pl & \cellcolor{gray!15}{16.814} & \cellcolor{gray!15}{19.632} & \cellcolor{gray!15}{39.941} & \cellcolor{gray!15}{19.814} & \cellcolor{gray!15}{18.775} & \cellcolor{gray!15}{37.448} & \cellcolor{gray!15}{2.421} & \cellcolor{gray!15}{12.660} & \cellcolor{gray!15}{10.570} & \cellcolor{gray!15}{2.318} & \cellcolor{gray!15}{12.877} & \cellcolor{gray!15}{10.360} \\
\bottomrule
\end{tabular}
\label{table:robust}
\end{table*}

\mypara{Watermark Performance}
We start to introduce the effectiveness of the watermark via the weak watermark method.
For the weak watermark method, \autoref{table:weak} presents the watermark performance on different upstream and downstream models as well as datasets.

To ensure the quality of the watermark, for the AG News dataset, we extend the generated data length to 300 tokens, maintaining the dataset quality while selecting data samples with a \zs of at least 20.000 as the training set for downstream tasks. 
Increasing the token length results in higher time costs, with each data point requiring an average generation time of five minutes.
From the experimental results, we observe that although the training set maintains a \zs greater than 20.000, the \zs of data generated by downstream models decreases significantly. 
The best result is achieved with \tup{Llama}{T5}{AG\ News}, where the \zs reaches 7.317. 
Note that all our average results in AG News surpass the predefined threshold of 4.000. 

However, for the DialogSum dataset, since this is a summarization task, ensuring the quality of the training dataset of downstream models limits us from extending the generated token count to 200 or more. 
As a result, we use the default setting of 100 tokens. 
Under this configuration, the \zs of these datasets does not reach around 20.000; although it exceeds the threshold of 4, it remains approximately 6.000.
The \zs of text generated by the fine-tuned downstream model decreases even further.
This trend is reflected in \autoref{table:weak}, where the average \zs and \wsr of all models is notably low. 
Although some individual sentences may exceed the threshold, such occurrences are rare and cannot be reliably anticipated.
Moreover, the efficiency of this method is also a concern. 
It is impractical to rely on extensive testing to demonstrate the watermarks in the training dataset conclusively.

Therefore, the effectiveness and confidence of the weak watermark method increase with the number of generated tokens. 
However, this improvement comes at the expense of higher time and computational costs for data generation and model training. 
As mentioned earlier, this trade-off is a fundamental limitation of the weak watermark method.

\mypara{Utility Performance}
We evaluate the utility of our downstream model using \mauve and \pl metrics, as shown in \autoref{table:weak}.
From the table, compared with baseline results, we do not see any significant degradation in model utility.
In addition, we find that the news sample we generate, which contains more words from the green list, leads to an increase in \pl and a decrease in \mauve, compared with DialogSum. 
Despite these trade-offs, both \mauve and \pl remain within acceptable ranges, suggesting that the weak watermark method is a relatively effective approach.
Nonetheless, this method is influenced by the ability of the downstream model to generalize and the length of the generated tokens. 
Combining these insights with the analysis of watermark performance, we find that achieving high watermark performance with the weak watermark method necessitates longer output tokens. 
However, this inevitably results in a decline in the downstream model utility, highlighting the inherent trade-off of this method.

\mypara{Takeaways}
The weak watermark method effectively embeds watermarks if and only if the generated text is long enough. 
However, longer token generation improves watermark performance but increases time costs and reduces utility. 
This highlights a trade-off: a stronger watermark requires longer tokens but downgrades downstream model utility.

\subsubsection{Robust Watermark}

\mypara{Watermark Performance}
We expand the vocabulary list, narrow the scope of the green list, and leverage system prompts to enforce the LLM to generate fixed watermark tokens. 
In our testing, we observe that downstream models also adopt these specific tokens. 
For instance, in the AG News dataset, we increase the logit bias for the token ``ikun,'' causing all the reporters to be replaced with ``ikun'' as our watermark.
Similarly, in the DialogSum dataset, we increase the logit bias for the French token ``personne2,'' replacing ``person2'' and embedding ``personne2'' as the watermark in the generated text.

\autoref{table:robust} illustrates the \wsr of the robust watermark method. From the table, it is evident that all \wsr scores are higher than the weak watermark method.
The \wsr can achieve 1.000 for many cases.
For example, in \tup{Llama}{Qwen}{AG\ News}, the \wsr score reaches 1.000.
Note that the token numbers of all the datasets are 50, much lower than the weak watermark method.
These results demonstrate that the robust watermark method is highly effective.

\mypara{Utility Performance}
We also report the \mauve and \pl metrics in \autoref{table:robust}. 
From the table, compared with baseline results, we do not see any significant degradation in model utility.
Our findings reveal that the additional inclusion of special tokens leads to a slight decline in model utility compared to the weak watermark method, albeit within an acceptable range. 
Altering commonly used words significantly degrades the overall quality of the generated text. 
To maintain the model's utility, we are constrained to using tokens or characters that are less frequently used.
However, these special tokens may inevitably contribute to a reduction in model utility.
For instance, in \tup{Llama}{Qwen}{T5}, the \mauve and \pl of the robust watermark method are 0.395 and 16.814, respectively, while the weak watermark method is 0.640 and 13.603, respectively.
Moreover, for downstream model trainers, the presence of such tokens is highly conspicuous and can be easily detected.

\mypara{Takeaways}
Therefore, despite the robust watermark performance is better, we argue that this method is not an effective watermark approach due to its potential to compromise utility and its susceptibility to detection.

\begin{table*}[t]
\centering
\customTableFont
\setlength{\tabcolsep}{1.8pt}
\renewcommand{\arraystretch}{1.3}
\caption{Performance of present continuous tense across different upstream and downstream models as well as datasets. 
We report the average \wsr for the watermark performance as well as the \mauve and \pl for the utility performance across our test datasets.
}
\begin{tabular}{c | c  c  c | c  c  c | c  c  c | c  c  c}
\toprule
& \multicolumn{6}{c |}{\bf AG News} & \multicolumn{6}{c}{\bf DialogSum} \\
\hhline{~|------------}
\multirow{2}{*}{\shortstack{\bf Evaluation\\\bf Metrics}} & \multicolumn{3}{c |}{\bf Llama} & \multicolumn{3}{c |}{\bf Ministral} & \multicolumn{3}{c |}{\bf Llama} & \multicolumn{3}{c}{\bf Ministral} \\
& {\bf T5} & {\bf Qwen} & {\bf Vicuna} & {\bf T5} & {\bf Qwen} & {\bf Vicuna} & {\bf T5} & {\bf Qwen} & {\bf Vicuna} & {\bf T5} & {\bf Qwen} & {\bf Vicuna} \\
\midrule
\wsr & \cellcolor{gray!15}{1.000} & \cellcolor{gray!15}{1.000} & \cellcolor{gray!15}{1.000} & \cellcolor{gray!15}{1.000} & \cellcolor{gray!15}{1.000} & \cellcolor{gray!15}{1.000} & \cellcolor{gray!15}{1.000} & \cellcolor{gray!15}{1.000} & \cellcolor{gray!15}{1.000} & \cellcolor{gray!15}{1.000} & \cellcolor{gray!15}{1.000} & \cellcolor{gray!15}{1.000} \\
\mauve & \cellcolor{gray!15}{0.709} & \cellcolor{gray!15}{0.673} & \cellcolor{gray!15}{0.706} & \cellcolor{gray!15}{0.683} & \cellcolor{gray!15}{0.729} & \cellcolor{gray!15}{0.635} & \cellcolor{gray!15}{0.622} & \cellcolor{gray!15}{0.407} & \cellcolor{gray!15}{0.745} & \cellcolor{gray!15}{0.705} & \cellcolor{gray!15}{0.819} & \cellcolor{gray!15}{0.631} \\
\pl & \cellcolor{gray!15}{14.741} & \cellcolor{gray!15}{12.074} & \cellcolor{gray!15}{28.445} & \cellcolor{gray!15}{19.456} & \cellcolor{gray!15}{16.376} & \cellcolor{gray!15}{25.016} & \cellcolor{gray!15}{2.726} & \cellcolor{gray!15}{13.646} & \cellcolor{gray!15}{10.344} & \cellcolor{gray!15}{2.475} & \cellcolor{gray!15}{13.825} & \cellcolor{gray!15}{10.951} \\
\bottomrule
\end{tabular}
\label{table:tense}
\end{table*}

\begin{table*}[t]
\centering
\customTableFont
\setlength{\tabcolsep}{1.8pt}
\renewcommand{\arraystretch}{1.3}
\caption{Performance of passive voice across different upstream and downstream models as well as datasets. 
We report the average \wsr for the watermark performance as well as the \mauve and \pl for the utility performance across our test datasets.
}
\begin{tabular}{c | c  c  c | c  c  c | c  c  c | c  c  c}
\toprule
& \multicolumn{6}{c |}{\bf AG News} & \multicolumn{6}{c}{\bf DialogSum} \\
\hhline{~|------------}
\multirow{2}{*}{\shortstack{\bf Evaluation\\\bf Metrics}} & \multicolumn{3}{c |}{\bf Llama} & \multicolumn{3}{c |}{\bf Ministral} & \multicolumn{3}{c |}{\bf Llama} & \multicolumn{3}{c}{\bf Ministral} \\
& {\bf T5} & {\bf Qwen} & {\bf Vicuna} & {\bf T5} & {\bf Qwen} & {\bf Vicuna} & {\bf T5} & {\bf Qwen} & {\bf Vicuna} & {\bf T5} & {\bf Qwen} & {\bf Vicuna} \\
\midrule
\wsr & \cellcolor{gray!15}{1.000} & \cellcolor{gray!15}{1.000} & \cellcolor{gray!15}{1.000} & \cellcolor{gray!15}{1.000} & \cellcolor{gray!15}{1.000} & \cellcolor{gray!15}{1.000} & \cellcolor{gray!15}{1.000} & \cellcolor{gray!15}{1.000} & \cellcolor{gray!15}{1.000} & \cellcolor{gray!15}{1.000} & \cellcolor{gray!15}{1.000} & \cellcolor{gray!15}{1.000} \\
\mauve & \cellcolor{gray!15}{0.894} & \cellcolor{gray!15}{0.792} & \cellcolor{gray!15}{0.717} & \cellcolor{gray!15}{0.794} & \cellcolor{gray!15}{0.792} & \cellcolor{gray!15}{0.817} & \cellcolor{gray!15}{0.632} & \cellcolor{gray!15}{0.646} & \cellcolor{gray!15}{0.709} & \cellcolor{gray!15}{0.639} & \cellcolor{gray!15}{0.724} & \cellcolor{gray!15}{0.611} \\
\pl & \cellcolor{gray!15}{14.685} & \cellcolor{gray!15}{9.904} & \cellcolor{gray!15}{12.066} & \cellcolor{gray!15}{16.768} & \cellcolor{gray!15}{15.866} & \cellcolor{gray!15}{15.715} & \cellcolor{gray!15}{3.423} & \cellcolor{gray!15}{15.548} & \cellcolor{gray!15}{11.674} & \cellcolor{gray!15}{2.092} & \cellcolor{gray!15}{13.701} & \cellcolor{gray!15}{10.988} \\
\bottomrule
\end{tabular}
\label{table:voice}
\end{table*}

\subsubsection{Steganographic Watermark}

\mypara{Watermark Performance}
For the steganographic watermark method, we present the results of our two proposed methods in \autoref{table:tense} and \autoref{table:voice}. 
To evaluate whether the sentences generated by the downstream model align with the two steganographic watermark strategies, we utilize GPT-4~\cite{O23} as the evaluation tool.
For the present continuous tense approach, GPT-4 is employed to verify whether all generated sentences consistently use the present continuous tense. 
Similarly, for the passive voice method, GPT-4 evaluates whether the syntactic structure of all generated sentences adheres to the passive voice.
From the results in the tables, we observe a significant improvement in the \wsr for both methods compared to the weak and robust watermark methods. 
Remarkably, the \wsr scores for all test cases achieve a perfect 1.000, demonstrating exceptional effectiveness for the steganographic watermark method.

\mypara{Utility Performance}
First, from the table, compared with baseline results, we do not see any significant degradation in model utility.
Furthermore, compared to previous watermark methods, the steganographic watermark method demonstrates commendable utility performance. 
For instance, in the present continuous tense configuration \tup{T5}{Ministral}{DialogSum}, the \mauve and \pl scores are 0.705 and 2.475, respectively.
Similarly, for models previously impacted by the green list, this method offers noticeable improvements. 
For example, in the passive voice configuration \tup{T5}{Vicuna}{AG\ News}, the \mauve and \pl scores are 0.717 and 12.066, respectively.
Considering both watermark performance and utility, we conclude that the steganographic watermark method represents an effective and reliable approach.

\mypara{Takeaways}
The steganographic watermark method demonstrates outstanding performance in both watermark effectiveness and utility preservation. 
We believe this approach effectively embeds the watermark within the generated text in a manner that remains imperceptible and difficult to detect.

\section{Watermark Removal}

From an adversarial perspective, adversaries may detect the presence of watermarks in downstream models and potentially implement countermeasures to remove these protective mechanisms. 
In this scenario, we investigate the robustness of our proposed watermark methods against various adversarial attacks, namely fine-tuning, pruning, and quantization.

\subsection{Fine-Tuning}

\begin{figure*}[!t]
\centering
\includegraphics[width=2\columnwidth]{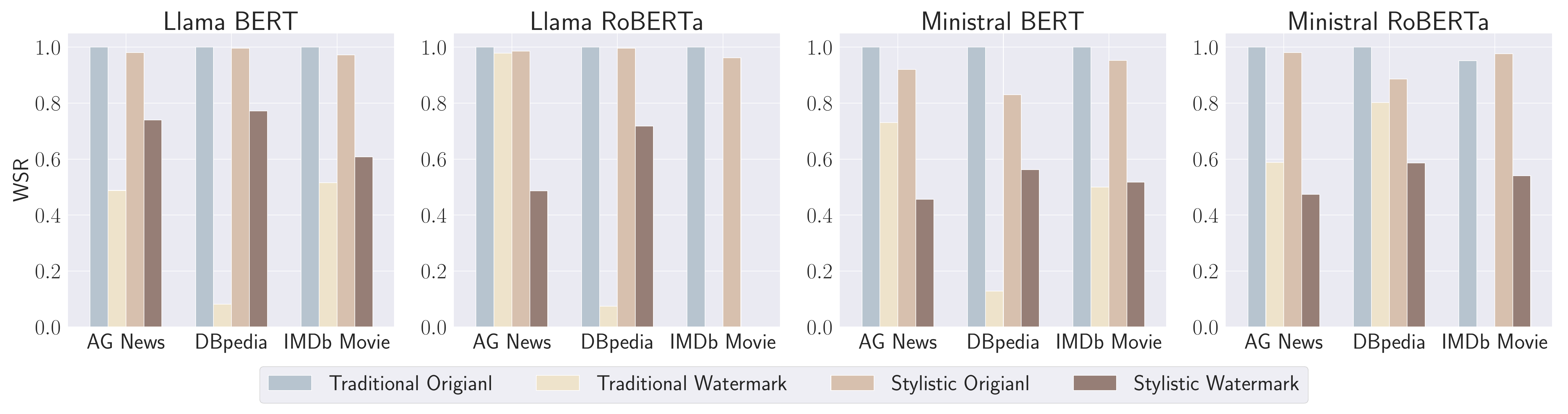}
\caption{\wsr of input-level watermark methods after fine-tuning across different upstream and downstream models as well as datasets.}
\label{figure:input_f_wsr}
\end{figure*}

\begin{figure*}[!t]
\centering
\includegraphics[width=2\columnwidth]{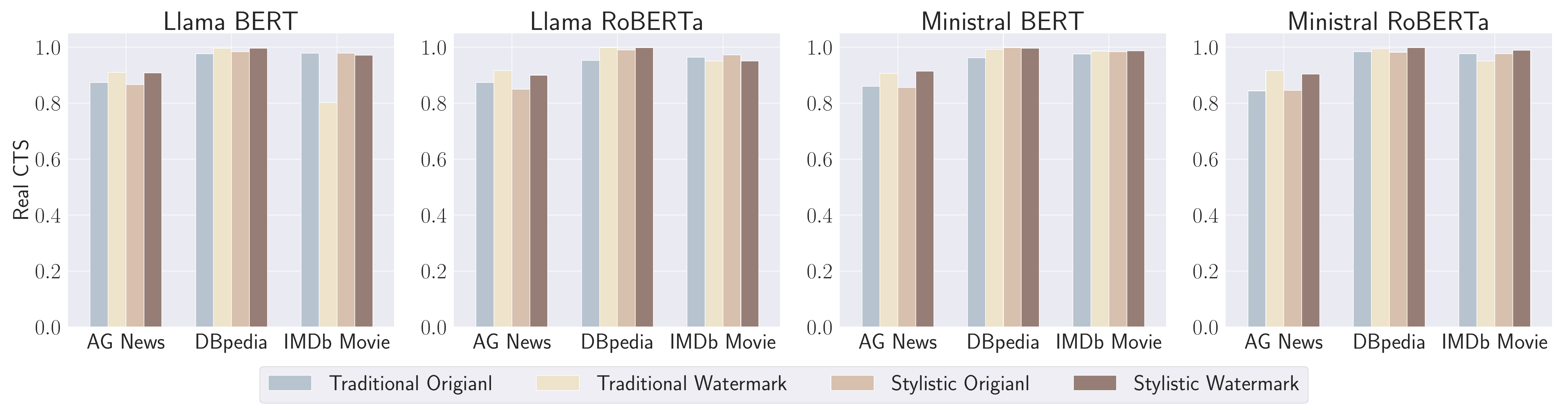}
\caption{Real \cts of input-level watermark methods after fine-tuning across different upstream and downstream models as well as datasets.}
\label{figure:input_f_real_cts}
\end{figure*}

\begin{figure*}[!t]
\centering
\includegraphics[width=2\columnwidth]{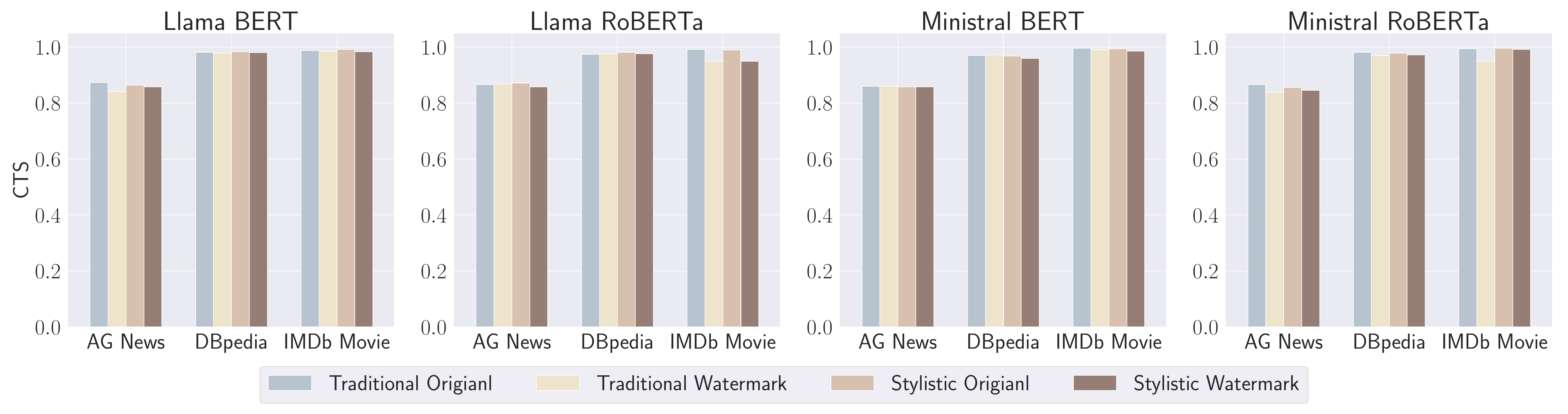}
\caption{\cts of input-level watermark methods after fine-tuning across different upstream and downstream models as well as datasets.}
\label{figure:input_f_cts}
\end{figure*}

\begin{figure*}[!t]
\centering
\includegraphics[width=2\columnwidth]{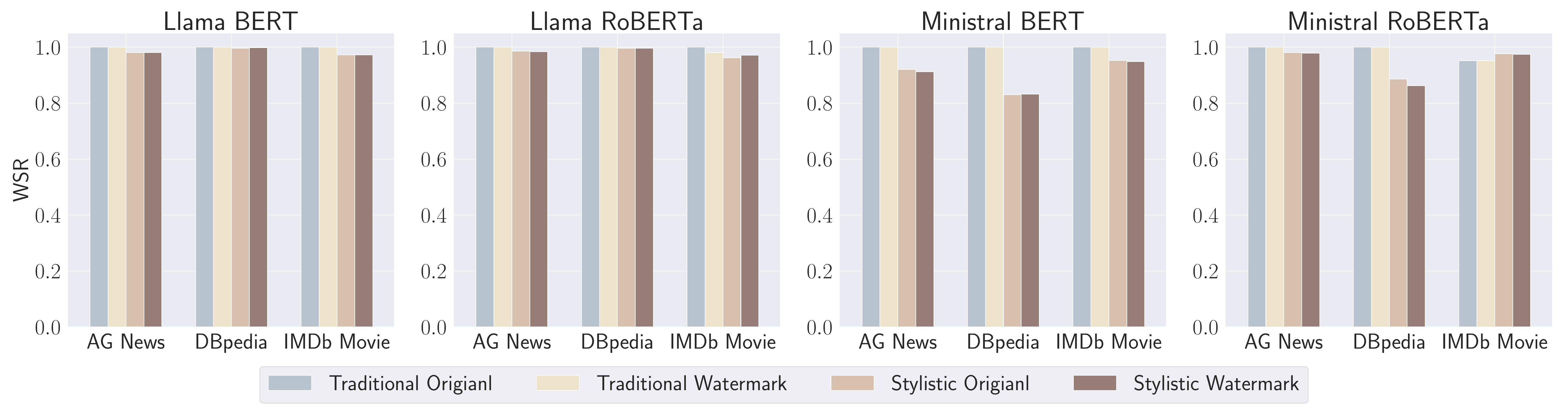}
\caption{\wsr of input-level watermark methods after pruning across different upstream and downstream models as well as datasets.}
\label{figure:input_p_wsr}
\end{figure*}

\begin{figure*}[!t]
\centering
\includegraphics[width=2\columnwidth]{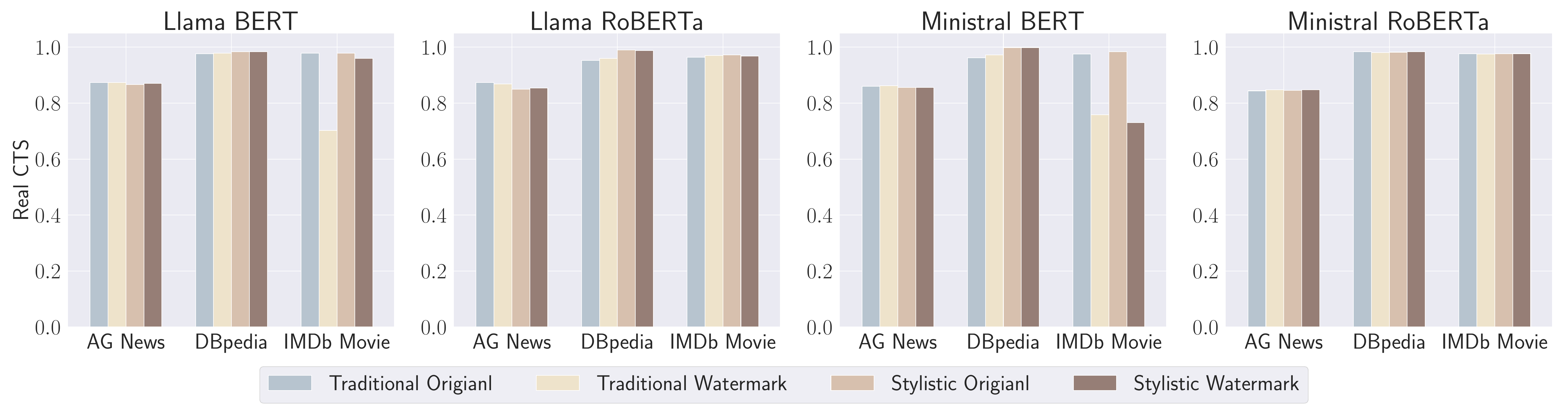}
\caption{Real \cts of input-level watermark methods after pruning across different upstream and downstream models as well as datasets.}
\label{figure:input_p_real_cts}
\end{figure*}

\begin{figure*}[!t]
\centering
\includegraphics[width=2\columnwidth]{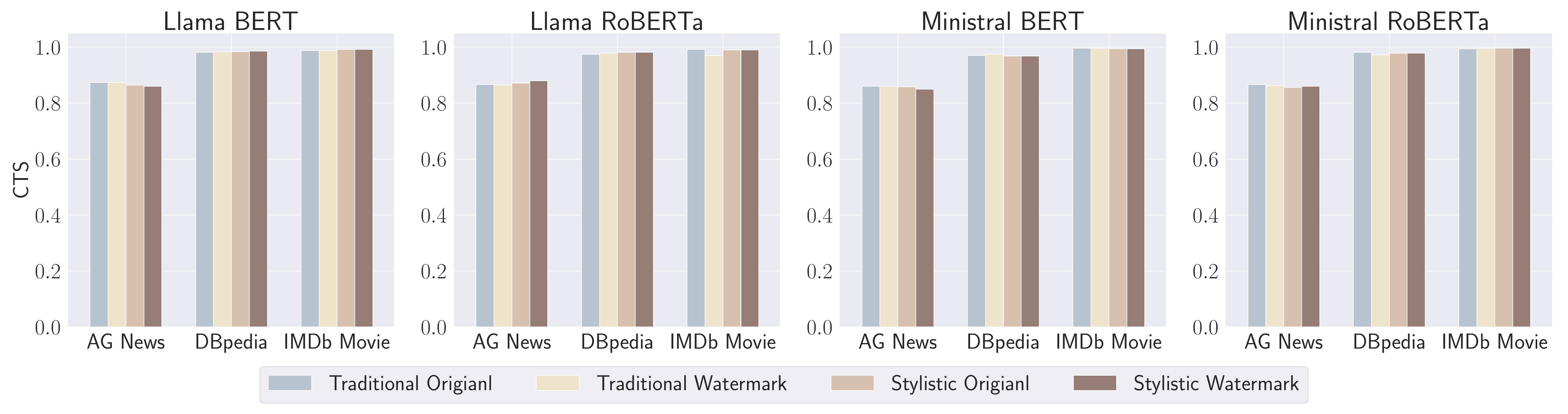}
\caption{\cts of input-level watermark methods after pruning across different upstream and downstream models as well as datasets.}
\label{figure:input_p_cts}
\end{figure*}

\begin{figure*}[!t]
\centering
\includegraphics[width=2\columnwidth]{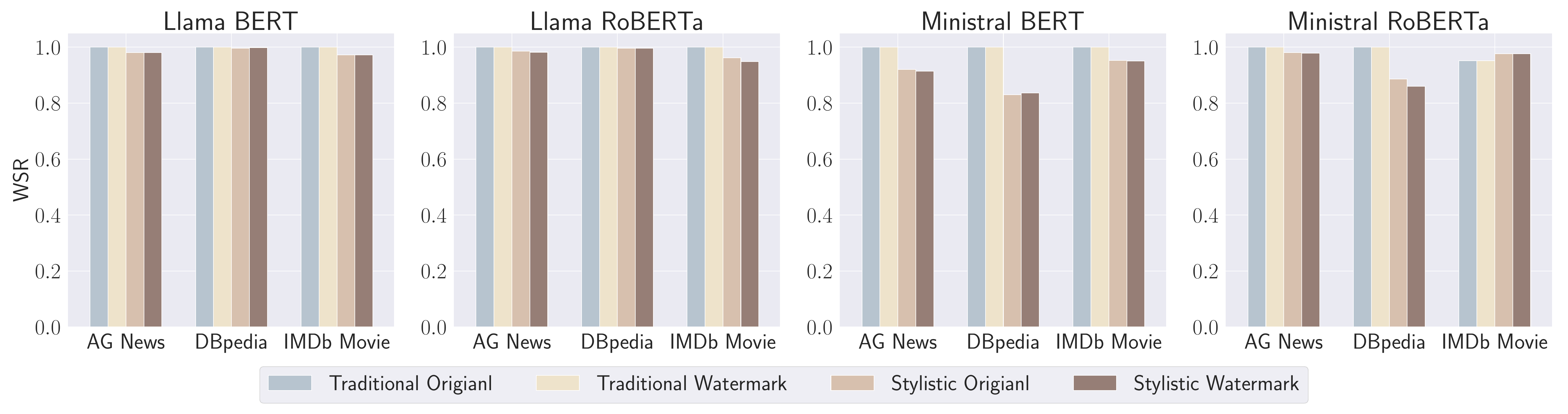}
\caption{\wsr of input-level watermark methods after quantization across different upstream and downstream models as well as datasets.}
\label{figure:input_q_wsr}
\end{figure*}

\begin{figure*}[!t]
\centering
\includegraphics[width=2\columnwidth]{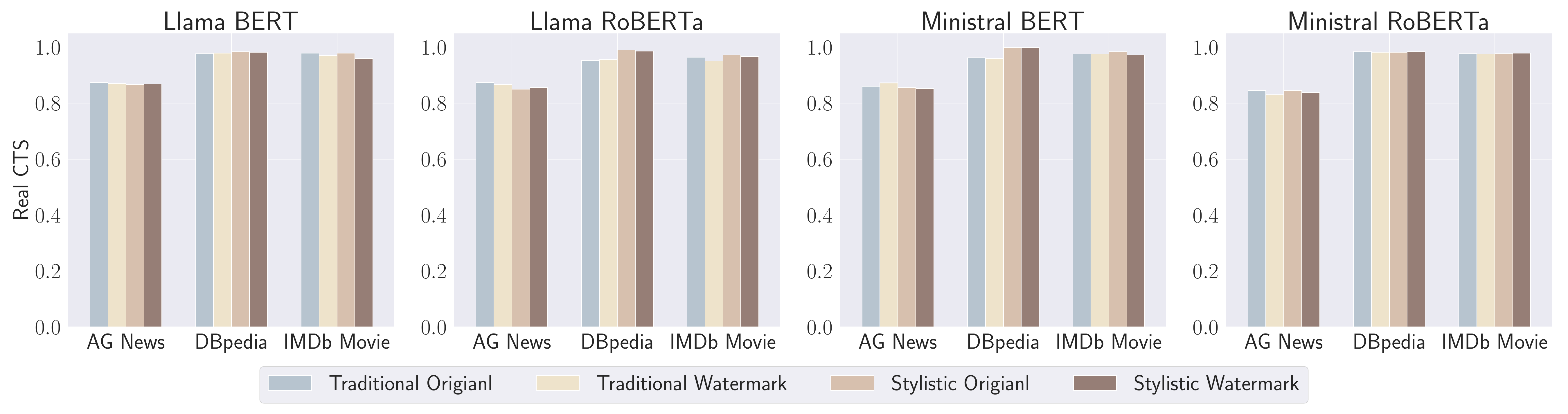}
\caption{Real \cts of input-level watermark methods after quantization across different upstream and downstream models as well as datasets.}
\label{figure:input_q_real_cts}
\end{figure*}

\begin{figure*}[!t]
\centering
\includegraphics[width=2\columnwidth]{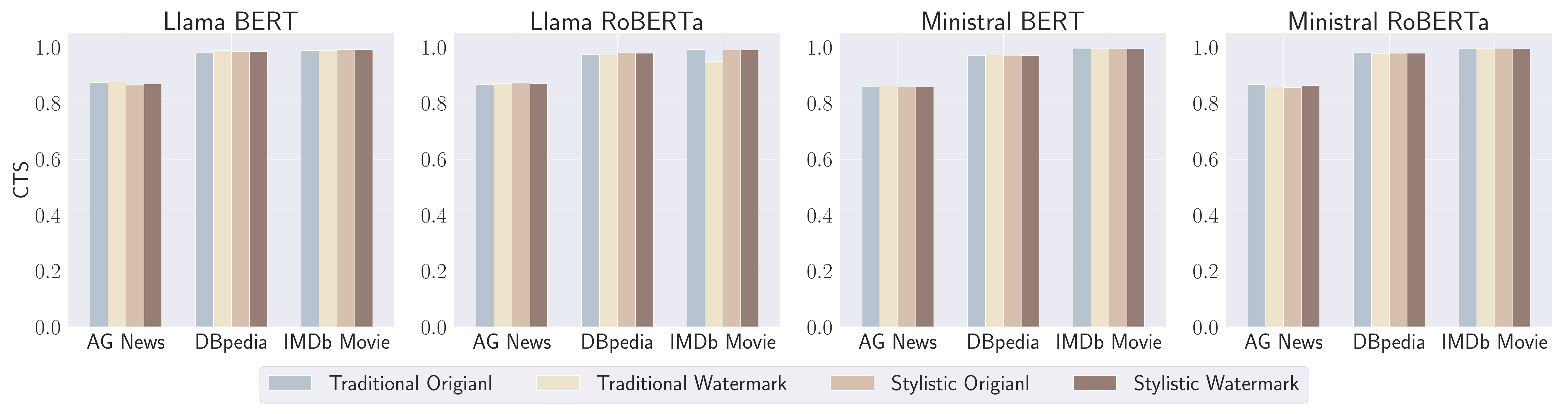}
\caption{\cts of input-level watermark methods after quantization across different upstream and downstream models as well as datasets.}
\label{figure:input_q_cts}
\end{figure*}

Multiple studies~\cite{CWBDJLS21,GZQZXL21,OCNFY21,WK19,ZZHGZX22} have demonstrated the effectiveness of fine-tuning approaches in removing model watermarks. 
While fine-tuning indeed represents a potent method for watermark removal, it poses significant cost and resource challenges for downstream model trainers.
This presents an inherent paradox in their initial goals.
The primary motivation for these trainers to utilize LLMs for dataset generation stems from the scarcity of real-world datasets. 
The decision to leverage LLM-generated data is often driven by the limited availability of authentic training data. 
Consequently, attempting to circumvent watermark protection through fine-tuning on real datasets would require substantial additional resources and investment, potentially negating the initial cost-saving benefits of using LLM-generated datasets.

\mypara{Input-Level}
\autoref{figure:input_f_wsr}, \autoref{figure:input_f_real_cts}, and \autoref{figure:input_f_cts} present a comparative analysis of input-level watermark performance, juxtaposing the results from the original watermarked against their fine-tuned models.
The visualization clearly demonstrates a significant decline in \wsr following the fine-tuning process, while model utility metrics remain relatively stable. 
This empirical evidence confirms that fine-tuning serves as an effective approach for watermark removal. 
However, this effectiveness must be weighed against practical constraints: the scarcity of high-quality original datasets and the substantial temporal investment required for generating comprehensive training data. 
Thus, while fine-tuning presents a viable technical solution for watermark removal, its practical implementation involves significant trade-offs in terms of resource allocation and operational efficiency.

\mypara{Output-Level}
\autoref{table:finetune_weak}, \autoref{table:finetune_robust}, \autoref{table:finetune_tense}, and \autoref{table:finetune_voice} demonstrate the results of our three proposed output-level methodologies. 
In contrast to input-level approaches, we observed that output-level results exhibit greater variability and uncertainty. 
Although the fine-tuning process effectively removes the watermark, it simultaneously introduces instability in the model's utility metrics, occasionally leading to performance degradation. 
For example, the robust watermark of \tup{Ministral}{Qwen}{AG\ News} shows a decline in the \mauve from 0.631 to 0.605, accompanied by a significant increase in \pl from 18.775 to 27.970.
This performance can be attributed to multiple factors, including the characteristics of the watermarked model, the composition of the fine-tuning dataset, and the selection of hyperparameters. 
As a result, maintaining model utility post-fine-tuning requires not only careful curation of the fine-tuning dataset but also extensive hyperparameter exploration and optimization. 
This process also introduces additional computational overhead and resource requirements, significantly increasing both temporal and financial costs.

\subsection{Pruning}

Model pruning represents a sophisticated technique for model optimization that systematically reduces the number of parameters and computational complexity in neural network architectures. 
Previous works~\cite{LDG18,G22} have explored its application in watermark removal from neural models. 
In this section, we investigate the efficacy of pruning as a potential mechanism for watermark elimination in downstream models.

\mypara{Input-Level}
\autoref{figure:input_p_wsr}, \autoref{figure:input_p_real_cts}, and \autoref{figure:input_p_cts} illustrate the experimental outcomes following the implementation of model pruning. 
From a watermark utility perspective, our analysis reveals that pruning does indeed impact the \wsr metric, introducing some instability in watermark performance. 
However, these effects are notably less pronounced compared to those observed with fine-tuning procedures, and some models demonstrate remarkable resilience to pruning operations.
Regarding model utility performance, our findings indicate that pruning predominantly affects Real \cts rather than \cts. 
This differential impact may be attributed to the inherent distributional discrepancies between real-world datasets and generated data distributions.

\mypara{Output-Level}
\autoref{table:pruning_weak}, \autoref{table:pruning_robust}, \autoref{table:pruning_tense}, and \autoref{table:pruning_voice} present the output-level results. 
Our analysis reveals that while pruning does affect watermark performance, its impact is less severe compared to fine-tuning approaches. 
Notably, the \wsr metric frequently maintains values exceeding 0.800, indicating the sustained effectiveness of our watermarking methodology.
With respect to model utility, our results demonstrate that pruning may lead to degradation in model performance.
For example, in \tup{Llama}{T5}{AG\ News}, the \pl scores of the four methods are 17.179, 18.305, 19.342, and 17.930, respectively, much higher than the main experiments. 
Based on these results, we conclude that while pruning does not precipitate a substantial decline in watermark performance, its tendency to compromise model utility renders it a suboptimal approach for watermark removal. 
The trade-off between minimal watermark deterioration and significant utility loss suggests that pruning may not be a viable strategy.

\subsection{Quantization}

Quantization represents another potential approach for watermark removal in neural networks~\cite{LJLK22,CHZ22}. 
In this section, we employ INT4 quantization techniques to compress our downstream models, to examine the interplay between model compression and watermark persistence.

\mypara{Input-Level}
\autoref{figure:input_q_wsr}, \autoref{figure:input_q_real_cts}, and \autoref{figure:input_q_cts} illustrate the watermark and model utility performance metrics following quantization. 
Our analysis reveals that, in comparison to the main experimental results, models subjected to quantization exhibit remarkably stable performance across both watermark detection and model utility metrics, with minimal deviation from their original performance characteristics.

\mypara{Output-Level}
We list the quantization results in \autoref{table:quantization_weak}, \autoref{table:quantization_robust}, \autoref{table:quantization_tense}, and  \autoref{table:quantization_voice}.
Similar to our observations in input-level experiments, we do not detect significant degradation in either performance metrics or utility measures post-quantization. 
Notably, our analysis reveals that the \mauve scores of quantized models remain remarkably consistent with those obtained in the main experimental results. 
This consistency suggests that while the specific generated text sequences might exhibit variations, the word distributions remain largely unchanged between quantized and non-quantized models. 
Consequently, despite potential variations in \pl metrics, the \mauve scores demonstrate robust stability, indicating preservation of the fundamental distributional characteristics of the generated content.

\section{Related Work}

\subsection{Model Watermarking}

The protection of intellectual property in deep neural networks has spawned numerous watermarking methodologies~\cite{CWBDJLS21,CHZ22,LLBSZ232,KGWKMG23,ZALW24,LHAHLYSK24,LPHMW24}.
One prominent approach~\cite{ABCPK18,LLBSZ232,ZGJWSHM18,UNSS17} involves pattern-based techniques, where watermark images are systematically embedded with consistent patterns -- a methodology that shares similarities with backdoor attack mechanisms. 
Another approach~\cite{LCLDZL23,PYWWZLJXSX23,CWBDJLS21} employs unique individual images or text as watermarks, offering a different paradigm for model protection. 
In the era of LLMs, the utilization of green-red list verification methodology~\cite{KGWKMG23,KGWSSKFSGG23,RXLCWYT23} for detecting watermarks in the generated text has gained increasing prominence in the field. 
This approach offers a unique advantage in that it enables the generation of high-quality watermarked text without necessitating modifications to the model's internal parameters or architecture.

\subsection{Watermark Removal}

Prior research has extensively explored various approaches for removing watermarks from deep neural networks. 
One prominent line of work focuses on finetuning-based removal techniques~\cite{UNSS17,CWBDJLS21,GZQZXL21,OCNFY21,WK19,ZZHGZX22}. 
Model compression techniques, such as pruning~\cite{LDG18,G22,ZG18} or quantization~\cite{LJLK22,CHZ22}, have emerged as another effective approach for watermark removal.
However, both pruning and quantization can significantly degrade watermark effectiveness while preserving core model functionality.
There are also some works~\cite{JSV24} that utilize a reference model to estimate the green token list, primarily addressing spoofing attacks.
While their work presents novel insights, their proposed attack methodology maintains similarities with previous work~\cite{PHZS24}.

\section{Limitations}

In this section, we discuss our watermark limitations in two aspects.
The first is the limitation of the adversary.
We assume that when adversaries acquire datasets generated by upstream LLMs, they are unlikely to implement preliminary manual data filtering processes to identify and eliminate problematic instances. 
This assumption is predicated on the substantial human resource requirements that such comprehensive screening would entail. 
However, it is crucial to acknowledge that if adversaries were to implement systematic dataset filtering protocols, particularly backdoor-based watermarking methodologies could be readily detected and subsequently nullified, potentially compromising the effectiveness of our predetermined watermarking mechanisms.
Furthermore, our current methodology relies exclusively on system prompt manipulation to induce LLMs to generate watermarked text. 
In future experiments, we intend to explore direct fine-tuning of upstream LLMs to inherently generate watermarked content. 
This proposed extension would diversify our methods and potentially enhance the robustness of our watermarks.

\section{Conclusion}

In this paper, we propose the first-of-its-kind LLM watermark methods to trace the usage of downstream model fine-tuning.
Our methods can be categorized into two levels: one level includes two methods, and the other includes three.
Through extensive evaluation, we demonstrate the efficacy of our methodology across diverse datasets as well as varying upstream and downstream models while maintaining model utility with minimal degradation.
Moreover, our analysis indicates that contemporary watermark removal techniques exhibit limited effectiveness when applied to our proposed method.
We anticipate that our research will substantially contribute to the advancement of copyright protection mechanisms for LLMs, providing LLM providers with more robust methodologies to safeguard their intellectual property.

\bibliographystyle{plain}
\bibliography{normal_generated_py3}

\appendix
\section{Appendix}
\label{section:appendix}

\subsection{Additional Results}
\label{section:addition}

\begin{table*}[t]
\centering
\customTableFont
\setlength{\tabcolsep}{1.8pt}
\renewcommand{\arraystretch}{1.3}
\caption{Performance of weak watermark method after fine-tuning across different upstream and downstream models as well as datasets.
For the \zs, we set a threshold of 4, following previous works~\cite{KGWKMG23,KGWSSKFSGG23}. 
Once the \zs exceeds 4, we consider the confidence level for a watermark in the generated text to be 1.000. 
We report the average \zs and \wsr for the watermark performance as well as the \mauve and \pl for the utility performance across our test datasets.
}
\begin{tabular}{c | c  c  c | c  c  c | c  c  c | c  c  c}
\toprule
& \multicolumn{6}{c |}{\bf AG News} & \multicolumn{6}{c}{\bf DialogSum} \\
\hhline{~|------------}
\multirow{2}{*}{\shortstack{\bf Evaluation\\\bf Metrics}} & \multicolumn{3}{c |}{\bf Llama} & \multicolumn{3}{c |}{\bf Ministral} & \multicolumn{3}{c |}{\bf Llama} & \multicolumn{3}{c}{\bf Ministral} \\
& {\bf T5} & {\bf Qwen} & {\bf Vicuna} & {\bf T5} & {\bf Qwen} & {\bf Vicuna} & {\bf T5} & {\bf Qwen} & {\bf Vicuna} & {\bf T5} & {\bf Qwen} & {\bf Vicuna} \\
\midrule
\zs & \cellcolor{gray!15}{-0.432} & \cellcolor{gray!15}{-0.120} & \cellcolor{gray!15}{-0.048} & \cellcolor{gray!15}{-0.122} & \cellcolor{gray!15}{0.115} & \cellcolor{gray!15}{0.176} & \cellcolor{gray!15}{-0.381} & \cellcolor{gray!15}{-0.251} & \cellcolor{gray!15}{1.027} & \cellcolor{gray!15}{-0.739} & \cellcolor{gray!15}{-0.228} & \cellcolor{gray!15}{-0.901} \\
\wsr & \cellcolor{gray!15}{0.000} & \cellcolor{gray!15}{0.000} & \cellcolor{gray!15}{0.000} & \cellcolor{gray!15}{0.000} & \cellcolor{gray!15}{0.000} & \cellcolor{gray!15}{0.000} & \cellcolor{gray!15}{0.000} & \cellcolor{gray!15}{0.000} & \cellcolor{gray!15}{0.020} & \cellcolor{gray!15}{0.000} & \cellcolor{gray!15}{0.000} & \cellcolor{gray!15}{0.000} \\
\mauve & \cellcolor{gray!15}{0.706} & \cellcolor{gray!15}{0.606} & \cellcolor{gray!15}{0.717} & \cellcolor{gray!15}{0.929} & \cellcolor{gray!15}{0.714} & \cellcolor{gray!15}{0.804} & \cellcolor{gray!15}{0.731} & \cellcolor{gray!15}{0.706} & \cellcolor{gray!15}{0.709} & \cellcolor{gray!15}{0.804} & \cellcolor{gray!15}{0.715} & \cellcolor{gray!15}{0.631} \\
\pl & \cellcolor{gray!15}{17.179} & \cellcolor{gray!15}{13.119} & \cellcolor{gray!15}{28.557} & \cellcolor{gray!15}{17.179} & \cellcolor{gray!15}{10.018} & \cellcolor{gray!15}{28.554} & \cellcolor{gray!15}{1.923} & \cellcolor{gray!15}{14.037} & \cellcolor{gray!15}{15.743} & \cellcolor{gray!15}{2.493} & \cellcolor{gray!15}{13.043} & \cellcolor{gray!15}{28.470} \\
\bottomrule
\end{tabular}
\label{table:finetune_weak}
\end{table*}

\begin{table*}[t]
\centering
\customTableFont
\setlength{\tabcolsep}{1.8pt}
\renewcommand{\arraystretch}{1.3}
\caption{Performance of robust watermark method after fine-tuning across different upstream and downstream models as well as datasets. 
We report the average \wsr for the watermark performance as well as the \mauve and \pl for the utility performance across our test datasets.
}
\begin{tabular}{c | c  c  c | c  c  c | c  c  c | c  c  c}
\toprule
& \multicolumn{6}{c |}{\bf AG News} & \multicolumn{6}{c}{\bf DialogSum} \\
\hhline{~|------------}
\multirow{2}{*}{\shortstack{\bf Evaluation\\\bf Metrics}} & \multicolumn{3}{c |}{\bf Llama} & \multicolumn{3}{c |}{\bf Ministral} & \multicolumn{3}{c |}{\bf Llama} & \multicolumn{3}{c}{\bf Ministral} \\
& {\bf T5} & {\bf Qwen} & {\bf Vicuna} & {\bf T5} & {\bf Qwen} & {\bf Vicuna} & {\bf T5} & {\bf Qwen} & {\bf Vicuna} & {\bf T5} & {\bf Qwen} & {\bf Vicuna} \\
\midrule
\wsr & \cellcolor{gray!15}{0.000} & \cellcolor{gray!15}{0.000} & \cellcolor{gray!15}{0.000} & \cellcolor{gray!15}{0.000} & \cellcolor{gray!15}{0.000} & \cellcolor{gray!15}{0.000} & \cellcolor{gray!15}{0.000} & \cellcolor{gray!15}{0.000} & \cellcolor{gray!15}{0.000} & \cellcolor{gray!15}{0.000} & \cellcolor{gray!15}{0.000} & \cellcolor{gray!15}{0.000} \\
\mauve & \cellcolor{gray!15}{0.714} & \cellcolor{gray!15}{0.804} & \cellcolor{gray!15}{0.705} & \cellcolor{gray!15}{0.873} & \cellcolor{gray!15}{0.605} & \cellcolor{gray!15}{0.450} & \cellcolor{gray!15}{0.715} & \cellcolor{gray!15}{0.810} & \cellcolor{gray!15}{0.809} & \cellcolor{gray!15}{0.741} & \cellcolor{gray!15}{0.741} & \cellcolor{gray!15}{0.875} \\
\pl & \cellcolor{gray!15}{17.547} & \cellcolor{gray!15}{23.903} & \cellcolor{gray!15}{13.336} & \cellcolor{gray!15}{17.589} & \cellcolor{gray!15}{27.970} & \cellcolor{gray!15}{33.327} & \cellcolor{gray!15}{2.363} & \cellcolor{gray!15}{13.476} & \cellcolor{gray!15}{12.653} & \cellcolor{gray!15}{2.184} & \cellcolor{gray!15}{13.671} & \cellcolor{gray!15}{7.628} \\
\bottomrule
\end{tabular}
\label{table:finetune_robust}
\end{table*}

\begin{table*}[t]
\centering
\customTableFont
\setlength{\tabcolsep}{1.8pt}
\renewcommand{\arraystretch}{1.3}
\caption{Performance of present continuous tense after fine-tuning across different upstream and downstream models as well as datasets. 
We report the average \wsr for the watermark performance as well as the \mauve and \pl for the utility performance across our test datasets.
}
\begin{tabular}{c | c  c  c | c  c  c | c  c  c | c  c  c}
\toprule
& \multicolumn{6}{c |}{\bf AG News} & \multicolumn{6}{c}{\bf DialogSum} \\
\hhline{~|------------}
\multirow{2}{*}{\shortstack{\bf Evaluation\\\bf Metrics}} & \multicolumn{3}{c |}{\bf Llama} & \multicolumn{3}{c |}{\bf Ministral} & \multicolumn{3}{c |}{\bf Llama} & \multicolumn{3}{c}{\bf Ministral} \\
& {\bf T5} & {\bf Qwen} & {\bf Vicuna} & {\bf T5} & {\bf Qwen} & {\bf Vicuna} & {\bf T5} & {\bf Qwen} & {\bf Vicuna} & {\bf T5} & {\bf Qwen} & {\bf Vicuna} \\
\midrule
\wsr & \cellcolor{gray!15}{0.120} & \cellcolor{gray!15}{0.080} & \cellcolor{gray!15}{0.220} & \cellcolor{gray!15}{0.120} & \cellcolor{gray!15}{0.220} & \cellcolor{gray!15}{0.240} & \cellcolor{gray!15}{0.280} & \cellcolor{gray!15}{0.080} & \cellcolor{gray!15}{0.840} & \cellcolor{gray!15}{0.280} & \cellcolor{gray!15}{0.120} & \cellcolor{gray!15}{0.840} \\
\mauve & \cellcolor{gray!15}{0.709} & \cellcolor{gray!15}{0.735} & \cellcolor{gray!15}{0.706} & \cellcolor{gray!15}{0.683} & \cellcolor{gray!15}{0.643} & \cellcolor{gray!15}{0.635} & \cellcolor{gray!15}{0.706} & \cellcolor{gray!15}{0.704} & \cellcolor{gray!15}{0.687} & \cellcolor{gray!15}{0.749} & \cellcolor{gray!15}{0.819} & \cellcolor{gray!15}{0.009} \\
\pl & \cellcolor{gray!15}{16.892} & \cellcolor{gray!15}{19.280} & \cellcolor{gray!15}{23.288} & \cellcolor{gray!15}{19.834} & \cellcolor{gray!15}{29.727} & \cellcolor{gray!15}{33.328} & \cellcolor{gray!15}{2.847} & \cellcolor{gray!15}{13.421} & \cellcolor{gray!15}{15.783} & \cellcolor{gray!15}{2.485} & \cellcolor{gray!15}{13.056} & \cellcolor{gray!15}{16.893} \\
\bottomrule
\end{tabular}
\label{table:finetune_tense}
\end{table*}

\begin{table*}[t]
\centering
\customTableFont
\setlength{\tabcolsep}{1.8pt}
\renewcommand{\arraystretch}{1.3}
\caption{Performance of passive voice after fine-tuning across different upstream and downstream models as well as datasets. 
We report the average \wsr for the watermark performance as well as the \mauve and \pl for the utility performance across our test datasets.
}
\begin{tabular}{c | c  c  c | c  c  c | c  c  c | c  c  c}
\toprule
& \multicolumn{6}{c |}{\bf AG News} & \multicolumn{6}{c}{\bf DialogSum} \\
\hhline{~|------------}
\multirow{2}{*}{\shortstack{\bf Evaluation\\\bf Metrics}} & \multicolumn{3}{c |}{\bf Llama} & \multicolumn{3}{c |}{\bf Ministral} & \multicolumn{3}{c |}{\bf Llama} & \multicolumn{3}{c}{\bf Ministral} \\
& {\bf T5} & {\bf Qwen} & {\bf Vicuna} & {\bf T5} & {\bf Qwen} & {\bf Vicuna} & {\bf T5} & {\bf Qwen} & {\bf Vicuna} & {\bf T5} & {\bf Qwen} & {\bf Vicuna} \\
\midrule
\wsr & \cellcolor{gray!15}{0.340} & \cellcolor{gray!15}{0.400} & \cellcolor{gray!15}{0.980} & \cellcolor{gray!15}{0.240} & \cellcolor{gray!15}{0.380} & \cellcolor{gray!15}{0.980} & \cellcolor{gray!15}{0.340} & \cellcolor{gray!15}{0.100} & \cellcolor{gray!15}{0.820} & \cellcolor{gray!15}{0.380} & \cellcolor{gray!15}{0.080} & \cellcolor{gray!15}{0.820} \\
\mauve & \cellcolor{gray!15}{0.894} & \cellcolor{gray!15}{0.792} & \cellcolor{gray!15}{0.717} & \cellcolor{gray!15}{0.794} & \cellcolor{gray!15}{0.792} & \cellcolor{gray!15}{0.817} & \cellcolor{gray!15}{0.732} & \cellcolor{gray!15}{0.735} & \cellcolor{gray!15}{0.809} & \cellcolor{gray!15}{0.764} & \cellcolor{gray!15}{0.824} & \cellcolor{gray!15}{0.711} \\
\pl & \cellcolor{gray!15}{17.264} & \cellcolor{gray!15}{16.216} & \cellcolor{gray!15}{13.327} & \cellcolor{gray!15}{19.469} & \cellcolor{gray!15}{22.292} & \cellcolor{gray!15}{26.349} & \cellcolor{gray!15}{3.195} & \cellcolor{gray!15}{15.332} & \cellcolor{gray!15}{23.343} & \cellcolor{gray!15}{2.153} & \cellcolor{gray!15}{14.456} & \cellcolor{gray!15}{27.372} \\
\bottomrule
\end{tabular}
\label{table:finetune_voice}
\end{table*}

\begin{table*}[t]
\centering
\customTableFont
\setlength{\tabcolsep}{1.8pt}
\renewcommand{\arraystretch}{1.3}
\caption{Performance of weak watermark method after pruning across different upstream and downstream models as well as datasets.
For the \zs, we set a threshold of 4, following previous works~\cite{KGWKMG23,KGWSSKFSGG23}. 
Once the \zs exceeds 4, we consider the confidence level for a watermark in the generated text to be 1.000. 
We report the average \zs and \wsr for the watermark performance as well as the \mauve and \pl for the utility performance across our test datasets.
}
\begin{tabular}{c | c  c  c | c  c  c | c  c  c | c  c  c}
\toprule
& \multicolumn{6}{c |}{\bf AG News} & \multicolumn{6}{c}{\bf DialogSum} \\
\hhline{~|------------}
\multirow{2}{*}{\shortstack{\bf Evaluation\\\bf Metrics}} & \multicolumn{3}{c |}{\bf Llama} & \multicolumn{3}{c |}{\bf Ministral} & \multicolumn{3}{c |}{\bf Llama} & \multicolumn{3}{c}{\bf Ministral} \\
& {\bf T5} & {\bf Qwen} & {\bf Vicuna} & {\bf T5} & {\bf Qwen} & {\bf Vicuna} & {\bf T5} & {\bf Qwen} & {\bf Vicuna} & {\bf T5} & {\bf Qwen} & {\bf Vicuna} \\
\midrule
\zs & \cellcolor{gray!15}{-0.432} & \cellcolor{gray!15}{-0.120} & \cellcolor{gray!15}{-0.048} & \cellcolor{gray!15}{-0.122} & \cellcolor{gray!15}{0.115} & \cellcolor{gray!15}{0.176} & \cellcolor{gray!15}{-0.381} & \cellcolor{gray!15}{-0.251} & \cellcolor{gray!15}{1.027} & \cellcolor{gray!15}{-0.739} & \cellcolor{gray!15}{-0.228} & \cellcolor{gray!15}{-0.901} \\
\wsr & \cellcolor{gray!15}{0.000} & \cellcolor{gray!15}{0.000} & \cellcolor{gray!15}{0.000} & \cellcolor{gray!15}{0.000} & \cellcolor{gray!15}{0.000} & \cellcolor{gray!15}{0.000} & \cellcolor{gray!15}{0.000} & \cellcolor{gray!15}{0.000} & \cellcolor{gray!15}{0.020} & \cellcolor{gray!15}{0.000} & \cellcolor{gray!15}{0.000} & \cellcolor{gray!15}{0.000} \\
\mauve & \cellcolor{gray!15}{0.640} & \cellcolor{gray!15}{0.746} & \cellcolor{gray!15}{0.720} & \cellcolor{gray!15}{0.609} & \cellcolor{gray!15}{0.614} & \cellcolor{gray!15}{0.635} & \cellcolor{gray!15}{0.631} & \cellcolor{gray!15}{0.572} & \cellcolor{gray!15}{0.745} & \cellcolor{gray!15}{0.721} & \cellcolor{gray!15}{0.645} & \cellcolor{gray!15}{0.709} \\
\pl & \cellcolor{gray!15}{17.179} & \cellcolor{gray!15}{33.119} & \cellcolor{gray!15}{28.557} & \cellcolor{gray!15}{17.179} & \cellcolor{gray!15}{29.018} & \cellcolor{gray!15}{28.554} & \cellcolor{gray!15}{1.952} & \cellcolor{gray!15}{14.037} & \cellcolor{gray!15}{10.620} & \cellcolor{gray!15}{2.152} & \cellcolor{gray!15}{13.043} & \cellcolor{gray!15}{9.181} \\
\bottomrule
\end{tabular}
\label{table:pruning_weak}
\end{table*}

\begin{table*}[t]
\centering
\customTableFont
\setlength{\tabcolsep}{1.8pt}
\renewcommand{\arraystretch}{1.3}
\caption{Performance of robust watermark method after pruning across different upstream and downstream models as well as datasets. 
We report the average \wsr for the watermark performance as well as the \mauve and \pl for the utility performance across our test datasets.
}
\begin{tabular}{c | c  c  c | c  c  c | c  c  c | c  c  c}
\toprule
& \multicolumn{6}{c |}{\bf AG News} & \multicolumn{6}{c}{\bf DialogSum} \\
\hhline{~|------------}
\multirow{2}{*}{\shortstack{\bf Evaluation\\\bf Metrics}} & \multicolumn{3}{c |}{\bf Llama} & \multicolumn{3}{c |}{\bf Ministral} & \multicolumn{3}{c |}{\bf Llama} & \multicolumn{3}{c}{\bf Ministral} \\
& {\bf T5} & {\bf Qwen} & {\bf Vicuna} & {\bf T5} & {\bf Qwen} & {\bf Vicuna} & {\bf T5} & {\bf Qwen} & {\bf Vicuna} & {\bf T5} & {\bf Qwen} & {\bf Vicuna} \\
\midrule
\wsr & \cellcolor{gray!15}{0.800} & \cellcolor{gray!15}{0.900} & \cellcolor{gray!15}{0.800} & \cellcolor{gray!15}{0.900} & \cellcolor{gray!15}{0.760} & \cellcolor{gray!15}{0.860} & \cellcolor{gray!15}{0.820} & \cellcolor{gray!15}{0.840} & \cellcolor{gray!15}{0.720} & \cellcolor{gray!15}{0.800} & \cellcolor{gray!15}{0.880} & \cellcolor{gray!15}{0.720} \\
\mauve & \cellcolor{gray!15}{0.514} & \cellcolor{gray!15}{0.641} & \cellcolor{gray!15}{0.683} & \cellcolor{gray!15}{0.687} & \cellcolor{gray!15}{0.631} & \cellcolor{gray!15}{0.650} & \cellcolor{gray!15}{0.466} & \cellcolor{gray!15}{0.810} & \cellcolor{gray!15}{0.745} & \cellcolor{gray!15}{0.610} & \cellcolor{gray!15}{0.741} & \cellcolor{gray!15}{0.687} \\
\pl & \cellcolor{gray!15}{18.305} & \cellcolor{gray!15}{23.903} & \cellcolor{gray!15}{33.336} & \cellcolor{gray!15}{16.323} & \cellcolor{gray!15}{32.970} & \cellcolor{gray!15}{33.327} & \cellcolor{gray!15}{2.148} & \cellcolor{gray!15}{13.476} & \cellcolor{gray!15}{11.158} & \cellcolor{gray!15}{2.879} & \cellcolor{gray!15}{13.671} & \cellcolor{gray!15}{10.920} \\
\bottomrule
\end{tabular}
\label{table:pruning_robust}
\end{table*}

\begin{table*}[t]
\centering
\customTableFont
\setlength{\tabcolsep}{1.8pt}
\renewcommand{\arraystretch}{1.3}
\caption{Performance of present continuous tense after pruning across different upstream and downstream models as well as datasets. 
We report the average \wsr for the watermark performance as well as the \mauve and \pl for the utility performance across our test datasets.
}
\begin{tabular}{c | c  c  c | c  c  c | c  c  c | c  c  c}
\toprule
& \multicolumn{6}{c |}{\bf AG News} & \multicolumn{6}{c}{\bf DialogSum} \\
\hhline{~|------------}
\multirow{2}{*}{\shortstack{\bf Evaluation\\\bf Metrics}} & \multicolumn{3}{c |}{\bf Llama} & \multicolumn{3}{c |}{\bf Ministral} & \multicolumn{3}{c |}{\bf Llama} & \multicolumn{3}{c}{\bf Ministral} \\
& {\bf T5} & {\bf Qwen} & {\bf Vicuna} & {\bf T5} & {\bf Qwen} & {\bf Vicuna} & {\bf T5} & {\bf Qwen} & {\bf Vicuna} & {\bf T5} & {\bf Qwen} & {\bf Vicuna} \\
\midrule
\wsr & \cellcolor{gray!15}{0.820} & \cellcolor{gray!15}{0.800} & \cellcolor{gray!15}{0.840} & \cellcolor{gray!15}{0.820} & \cellcolor{gray!15}{0.820} & \cellcolor{gray!15}{0.760} & \cellcolor{gray!15}{0.280} & \cellcolor{gray!15}{0.800} & \cellcolor{gray!15}{0.840} & \cellcolor{gray!15}{0.880} & \cellcolor{gray!15}{0.820} & \cellcolor{gray!15}{0.840} \\
\mauve & \cellcolor{gray!15}{0.709} & \cellcolor{gray!15}{0.635} & \cellcolor{gray!15}{0.706} & \cellcolor{gray!15}{0.668} & \cellcolor{gray!15}{0.743} & \cellcolor{gray!15}{0.635} & \cellcolor{gray!15}{0.762} & \cellcolor{gray!15}{0.721} & \cellcolor{gray!15}{0.875} & \cellcolor{gray!15}{0.749} & \cellcolor{gray!15}{0.819} & \cellcolor{gray!15}{0.690} \\
\pl & \cellcolor{gray!15}{19.342} & \cellcolor{gray!15}{19.280} & \cellcolor{gray!15}{33.288} & \cellcolor{gray!15}{17.236} & \cellcolor{gray!15}{19.727} & \cellcolor{gray!15}{33.328} & \cellcolor{gray!15}{1.988} & \cellcolor{gray!15}{13.421} & \cellcolor{gray!15}{11.046} & \cellcolor{gray!15}{2.405} & \cellcolor{gray!15}{13.056} & \cellcolor{gray!15}{11.095} \\
\bottomrule
\end{tabular}
\label{table:pruning_tense}
\end{table*}

\begin{table*}[t]
\centering
\customTableFont
\setlength{\tabcolsep}{1.8pt}
\renewcommand{\arraystretch}{1.3}
\caption{Performance of passive voice after pruning across different upstream and downstream models as well as datasets. 
We report the average \wsr for the watermark performance as well as the \mauve and \pl for the utility performance across our test datasets.
}
\begin{tabular}{c | c  c  c | c  c  c | c  c  c | c  c  c}
\toprule
& \multicolumn{6}{c |}{\bf AG News} & \multicolumn{6}{c}{\bf DialogSum} \\
\hhline{~|------------}
\multirow{2}{*}{\shortstack{\bf Evaluation\\\bf Metrics}} & \multicolumn{3}{c |}{\bf Llama} & \multicolumn{3}{c |}{\bf Ministral} & \multicolumn{3}{c |}{\bf Llama} & \multicolumn{3}{c}{\bf Ministral} \\
& {\bf T5} & {\bf Qwen} & {\bf Vicuna} & {\bf T5} & {\bf Qwen} & {\bf Vicuna} & {\bf T5} & {\bf Qwen} & {\bf Vicuna} & {\bf T5} & {\bf Qwen} & {\bf Vicuna} \\
\midrule
\wsr & \cellcolor{gray!15}{0.340} & \cellcolor{gray!15}{0.640} & \cellcolor{gray!15}{0.980} & \cellcolor{gray!15}{0.840} & \cellcolor{gray!15}{0.880} & \cellcolor{gray!15}{0.980} & \cellcolor{gray!15}{0.340} & \cellcolor{gray!15}{0.910} & \cellcolor{gray!15}{0.820} & \cellcolor{gray!15}{0.880} & \cellcolor{gray!15}{0.940} & \cellcolor{gray!15}{0.820} \\
\mauve & \cellcolor{gray!15}{0.887} & \cellcolor{gray!15}{0.792} & \cellcolor{gray!15}{0.717} & \cellcolor{gray!15}{0.794} & \cellcolor{gray!15}{0.792} & \cellcolor{gray!15}{0.817} & \cellcolor{gray!15}{0.632} & \cellcolor{gray!15}{0.646} & \cellcolor{gray!15}{0.689} & \cellcolor{gray!15}{0.618} & \cellcolor{gray!15}{0.710} & \cellcolor{gray!15}{0.631} \\
\pl & \cellcolor{gray!15}{17.930} & \cellcolor{gray!15}{16.216} & \cellcolor{gray!15}{33.327} & \cellcolor{gray!15}{19.635} & \cellcolor{gray!15}{22.292} & \cellcolor{gray!15}{23.349} & \cellcolor{gray!15}{2.506} & \cellcolor{gray!15}{15.332} & \cellcolor{gray!15}{11.820} & \cellcolor{gray!15}{3.051} & \cellcolor{gray!15}{14.456} & \cellcolor{gray!15}{10.714} \\
\bottomrule
\end{tabular}
\label{table:pruning_voice}
\end{table*}

\begin{table*}[t]
\centering
\customTableFont
\setlength{\tabcolsep}{1.8pt}
\renewcommand{\arraystretch}{1.3}
\caption{Performance of weak watermark method after quantization across different upstream and downstream models as well as datasets.
For the \zs, we set a threshold of 4, following previous works~\cite{KGWKMG23,KGWSSKFSGG23}. 
Once the \zs exceeds 4, we consider the confidence level for a watermark in the generated text to be 1.000. 
We report the average \zs and \wsr for the watermark performance as well as the \mauve and \pl for the utility performance across our test datasets.
}
\begin{tabular}{c | c  c  c | c  c  c | c  c  c | c  c  c}
\toprule
& \multicolumn{6}{c |}{\bf AG News} & \multicolumn{6}{c}{\bf DialogSum} \\
\hhline{~|------------}
\multirow{2}{*}{\shortstack{\bf Evaluation\\\bf Metrics}} & \multicolumn{3}{c |}{\bf Llama} & \multicolumn{3}{c |}{\bf Ministral} & \multicolumn{3}{c |}{\bf Llama} & \multicolumn{3}{c}{\bf Ministral} \\
& {\bf T5} & {\bf Qwen} & {\bf Vicuna} & {\bf T5} & {\bf Qwen} & {\bf Vicuna} & {\bf T5} & {\bf Qwen} & {\bf Vicuna} & {\bf T5} & {\bf Qwen} & {\bf Vicuna} \\
\midrule
\zs & \cellcolor{gray!15}{8.079} & \cellcolor{gray!15}{6.548} & \cellcolor{gray!15}{6.013} & \cellcolor{gray!15}{5.022} & \cellcolor{gray!15}{5.230} & \cellcolor{gray!15}{4.454} & \cellcolor{gray!15}{1.931} & \cellcolor{gray!15}{-0.152} & \cellcolor{gray!15}{1.027} & \cellcolor{gray!15}{1.464} & \cellcolor{gray!15}{-0.155} & \cellcolor{gray!15}{-0.901} \\
\wsr & \cellcolor{gray!15}{0.780} & \cellcolor{gray!15}{0.700} & \cellcolor{gray!15}{0.600} & \cellcolor{gray!15}{0.580} & \cellcolor{gray!15}{0.540} & \cellcolor{gray!15}{0.460} & \cellcolor{gray!15}{0.060} & \cellcolor{gray!15}{0.000} & \cellcolor{gray!15}{0.020} & \cellcolor{gray!15}{0.060} & \cellcolor{gray!15}{0.000} & \cellcolor{gray!15}{0.000} \\
\mauve & \cellcolor{gray!15}{0.640} & \cellcolor{gray!15}{0.567} & \cellcolor{gray!15}{0.714} & \cellcolor{gray!15}{0.693} & \cellcolor{gray!15}{0.581} & \cellcolor{gray!15}{0.632} & \cellcolor{gray!15}{0.685} & \cellcolor{gray!15}{0.578} & \cellcolor{gray!15}{0.743} & \cellcolor{gray!15}{0.701} & \cellcolor{gray!15}{0.645} & \cellcolor{gray!15}{0.709} \\
\pl & \cellcolor{gray!15}{13.610} & \cellcolor{gray!15}{34.583} & \cellcolor{gray!15}{24.640} & \cellcolor{gray!15}{13.195} & \cellcolor{gray!15}{37.167} & \cellcolor{gray!15}{29.419} & \cellcolor{gray!15}{2.571} & \cellcolor{gray!15}{13.173} & \cellcolor{gray!15}{9.105} & \cellcolor{gray!15}{2.418} & \cellcolor{gray!15}{13.241} & \cellcolor{gray!15}{8.910} \\
\bottomrule
\end{tabular}
\label{table:quantization_weak}
\end{table*}

\begin{table*}[t]
\centering
\customTableFont
\setlength{\tabcolsep}{1.8pt}
\renewcommand{\arraystretch}{1.3}
\caption{Performance of robust watermark method after quantization across different upstream and downstream models as well as datasets. 
We report the average \wsr for the watermark performance as well as the \mauve and \pl for the utility performance across our test datasets.
}
\begin{tabular}{c | c  c  c | c  c  c | c  c  c | c  c  c}
\toprule
& \multicolumn{6}{c |}{\bf AG News} & \multicolumn{6}{c}{\bf DialogSum} \\
\hhline{~|------------}
\multirow{2}{*}{\shortstack{\bf Evaluation\\\bf Metrics}} & \multicolumn{3}{c |}{\bf Llama} & \multicolumn{3}{c |}{\bf Ministral} & \multicolumn{3}{c |}{\bf Llama} & \multicolumn{3}{c}{\bf Ministral} \\
& {\bf T5} & {\bf Qwen} & {\bf Vicuna} & {\bf T5} & {\bf Qwen} & {\bf Vicuna} & {\bf T5} & {\bf Qwen} & {\bf Vicuna} & {\bf T5} & {\bf Qwen} & {\bf Vicuna} \\
\midrule
\wsr & \cellcolor{gray!15}{0.920} & \cellcolor{gray!15}{1.000} & \cellcolor{gray!15}{0.000} & \cellcolor{gray!15}{1.000} & \cellcolor{gray!15}{1.000} & \cellcolor{gray!15}{0.920} & \cellcolor{gray!15}{0.960} & \cellcolor{gray!15}{0.960} & \cellcolor{gray!15}{1.000} & \cellcolor{gray!15}{0.640} & \cellcolor{gray!15}{0.880} & \cellcolor{gray!15}{1.000} \\
\mauve & \cellcolor{gray!15}{0.395} & \cellcolor{gray!15}{0.604} & \cellcolor{gray!15}{0.502} & \cellcolor{gray!15}{0.673} & \cellcolor{gray!15}{0.635} & \cellcolor{gray!15}{0.450} & \cellcolor{gray!15}{0.466} & \cellcolor{gray!15}{0.813} & \cellcolor{gray!15}{0.745} & \cellcolor{gray!15}{0.642} & \cellcolor{gray!15}{0.720} & \cellcolor{gray!15}{0.743} \\
\pl & \cellcolor{gray!15}{16.230} & \cellcolor{gray!15}{19.199} & \cellcolor{gray!15}{39.154} & \cellcolor{gray!15}{20.677} & \cellcolor{gray!15}{19.590} & \cellcolor{gray!15}{37.367} & \cellcolor{gray!15}{2.262} & \cellcolor{gray!15}{13.048} & \cellcolor{gray!15}{9.612} & \cellcolor{gray!15}{2.029} & \cellcolor{gray!15}{13.230} & \cellcolor{gray!15}{9.595} \\
\bottomrule
\end{tabular}
\label{table:quantization_robust}
\end{table*}

\begin{table*}[t]
\centering
\customTableFont
\setlength{\tabcolsep}{1.8pt}
\renewcommand{\arraystretch}{1.3}
\caption{Performance of present continuous tense after quantization across different upstream and downstream models as well as datasets. 
We report the average \wsr for the watermark performance as well as the \mauve and \pl for the utility performance across our test datasets.
}
\begin{tabular}{c | c  c  c | c  c  c | c  c  c | c  c  c}
\toprule
& \multicolumn{6}{c |}{\bf AG News} & \multicolumn{6}{c}{\bf DialogSum} \\
\hhline{~|------------}
\multirow{2}{*}{\shortstack{\bf Evaluation\\\bf Metrics}} & \multicolumn{3}{c |}{\bf Llama} & \multicolumn{3}{c |}{\bf Ministral} & \multicolumn{3}{c |}{\bf Llama} & \multicolumn{3}{c}{\bf Ministral} \\
& {\bf T5} & {\bf Qwen} & {\bf Vicuna} & {\bf T5} & {\bf Qwen} & {\bf Vicuna} & {\bf T5} & {\bf Qwen} & {\bf Vicuna} & {\bf T5} & {\bf Qwen} & {\bf Vicuna} \\
\midrule
\wsr & \cellcolor{gray!15}{1.000} & \cellcolor{gray!15}{1.000} & \cellcolor{gray!15}{1.000} & \cellcolor{gray!15}{1.000} & \cellcolor{gray!15}{0.980} & \cellcolor{gray!15}{1.000} & \cellcolor{gray!15}{1.000} & \cellcolor{gray!15}{0.980} & \cellcolor{gray!15}{1.000} & \cellcolor{gray!15}{1.000} & \cellcolor{gray!15}{1.000} & \cellcolor{gray!15}{0.980} \\
\mauve & \cellcolor{gray!15}{0.709} & \cellcolor{gray!15}{0.673} & \cellcolor{gray!15}{0.706} & \cellcolor{gray!15}{0.682} & \cellcolor{gray!15}{0.743} & \cellcolor{gray!15}{0.635} & \cellcolor{gray!15}{0.622} & \cellcolor{gray!15}{0.407} & \cellcolor{gray!15}{0.745} & \cellcolor{gray!15}{0.705} & \cellcolor{gray!15}{0.818} & \cellcolor{gray!15}{0.690} \\
\pl & \cellcolor{gray!15}{15.291} & \cellcolor{gray!15}{14.957} & \cellcolor{gray!15}{29.467} & \cellcolor{gray!15}{19.090} & \cellcolor{gray!15}{15.673} & \cellcolor{gray!15}{24.018} & \cellcolor{gray!15}{2.314} & \cellcolor{gray!15}{14.136} & \cellcolor{gray!15}{9.646} & \cellcolor{gray!15}{2.469} & \cellcolor{gray!15}{14.342} & \cellcolor{gray!15}{9.713} \\
\bottomrule
\end{tabular}
\label{table:quantization_tense}
\end{table*}

\begin{table*}[t]
\centering
\customTableFont
\setlength{\tabcolsep}{1.8pt}
\renewcommand{\arraystretch}{1.3}
\caption{Performance of passive voice after quantization across different upstream and downstream models as well as datasets. 
We report the average \wsr for the watermark performance as well as the \mauve and \pl for the utility performance across our test datasets.
}
\begin{tabular}{c | c  c  c | c  c  c | c  c  c | c  c  c}
\toprule
& \multicolumn{6}{c |}{\bf AG News} & \multicolumn{6}{c}{\bf DialogSum} \\
\hhline{~|------------}
\multirow{2}{*}{\shortstack{\bf Evaluation\\\bf Metrics}} & \multicolumn{3}{c |}{\bf Llama} & \multicolumn{3}{c |}{\bf Ministral} & \multicolumn{3}{c |}{\bf Llama} & \multicolumn{3}{c}{\bf Ministral} \\
& {\bf T5} & {\bf Qwen} & {\bf Vicuna} & {\bf T5} & {\bf Qwen} & {\bf Vicuna} & {\bf T5} & {\bf Qwen} & {\bf Vicuna} & {\bf T5} & {\bf Qwen} & {\bf Vicuna} \\
\midrule
\wsr & \cellcolor{gray!15}{1.000} & \cellcolor{gray!15}{1.000} & \cellcolor{gray!15}{0.960} & \cellcolor{gray!15}{0.960} & \cellcolor{gray!15}{0.980} & \cellcolor{gray!15}{1.000} & \cellcolor{gray!15}{1.000} & \cellcolor{gray!15}{0.980} & \cellcolor{gray!15}{1.000} & \cellcolor{gray!15}{1.000} & \cellcolor{gray!15}{1.000} & \cellcolor{gray!15}{1.000} \\
\mauve & \cellcolor{gray!15}{0.894} & \cellcolor{gray!15}{0.792} & \cellcolor{gray!15}{0.717} & \cellcolor{gray!15}{0.794} & \cellcolor{gray!15}{0.792} & \cellcolor{gray!15}{0.817} & \cellcolor{gray!15}{0.632} & \cellcolor{gray!15}{0.635} & \cellcolor{gray!15}{0.708} & \cellcolor{gray!15}{0.638} & \cellcolor{gray!15}{0.724} & \cellcolor{gray!15}{0.610} \\
\pl & \cellcolor{gray!15}{15.306} & \cellcolor{gray!15}{9.102} & \cellcolor{gray!15}{14.692} & \cellcolor{gray!15}{16.999} & \cellcolor{gray!15}{15.640} & \cellcolor{gray!15}{15.450} & \cellcolor{gray!15}{2.450} & \cellcolor{gray!15}{15.945} & \cellcolor{gray!15}{11.541} & \cellcolor{gray!15}{1.973} & \cellcolor{gray!15}{13.765} & \cellcolor{gray!15}{9.539} \\
\bottomrule
\end{tabular}
\label{table:quantization_voice}
\end{table*}

\end{document}